\documentclass[letter,11pt]{article}%
\usepackage{amsmath}
\usepackage{amsfonts}
\usepackage{amssymb}
\usepackage{booktabs}
\usepackage{float}
\usepackage{threeparttable}
\usepackage{multirow,longtable,bigstrut,caption}
\usepackage{subfig}
\usepackage{tabu}
\usepackage{tabularx,ragged2e,booktabs,caption}
\usepackage{pbox}
\usepackage[abs]{overpic}
\usepackage{graphicx}
\usepackage[usenames,dvipsnames]{color}
\usepackage{rotating}
\usepackage[height=9in,left=1in,right=0.75in,bottom=1in]{geometry}
\usepackage{booktabs}
\usepackage{longtable}
\usepackage[font=small,labelfont=bf]{caption}
\usepackage[breaklinks=true,bookmarksopen=true,colorlinks=true,citecolor=blue]%
{hyperref}
\usepackage[authoryear]{natbib}%
\providecommand{\U}[1]{\protect\rule{.1in}{.1in}}
\newtheorem{theorem}{Theorem}[section]

\newtheorem{lemma}[theorem]{Lemma}

\newtheorem{proposition}[theorem]{Proposition}
\newtheorem{remark}{Remark}[section]

\makeatletter

\@addtoreset{lemma}{Appendix B} \makeatother

\hypersetup{pdftitle={Escanciano, Li}, pdfsubject={On identification and estimation of structural linear functionals}, pdfauthor={Juan Carlos Escanciano, Wei Li}, pdfkeywords={Nonparametrics} }
\begin{document}

\title{Optimal Linear Instrumental Variables Approximations\thanks{We thank Andres
Santos, Pedro Sant'Anna, Oliver Linton, an AE, two anonymous referees and
seminar participants at Cambridge, LACEA-LAMES 2018, TSE, UCL and UC3M for
useful comments. }}
\author{Juan Carlos Escanciano\thanks{Department of Economics, Universidad Carlos III
de Madrid, 15.2.19, Calle Madrid 126, Getafe, Madrid, 28903, Spain. E-mail:
\href{mailto:jescanci@indiana.edu}{jescanci@econ.uc3m.es}. Web Page:
\href{https://sites.google.com/view/juancarlosescanciano}{https://sites.google.com/view/juancarlosescanciano}%
. Research funded by the Spanish Grant PGC2018-096732-B-I00.}\\\textit{Universidad Carlos III de Madrid}
\and Wei Li \thanks{Department of Mathematics, 215 Carnegie Building, Syracuse, New
York 13244, USA. E-mail: \href{mailto:wli35@ncsu.edu}{wli169@syr.edu} }\\\textit{Syracuse University}}
\date{February 3th, 2020}
\maketitle

\begin{abstract}
This paper studies the identification and estimation of the optimal linear
approximation of a structural regression function. The parameter in the linear
approximation is called the Optimal Linear Instrumental Variables
Approximation (OLIVA). This paper shows that a necessary condition for
standard inference on the OLIVA is also sufficient for the existence of an IV
estimand in a linear model. The instrument in the IV estimand is unknown and
may not be identified. A Two-Step IV (TSIV) estimator based on Tikhonov
regularization is proposed, which can be implemented by standard regression
routines. We establish the asymptotic normality of the TSIV estimator assuming
neither completeness nor identification of the instrument. As an important
application of our analysis, we robustify the classical Hausman test for
exogeneity against misspecification of the linear structural model. We also
discuss extensions to weighted least squares criteria. Monte Carlo simulations
suggest an excellent finite sample performance for the proposed inferences.
Finally, in an empirical application estimating the elasticity of
intertemporal substitution (EIS) with US data, we obtain TSIV estimates that
are much larger than their standard IV counterparts, with our robust Hausman
test failing to reject the null hypothesis of exogeneity of real interest rates.

\vspace{2mm}

\begin{description}
\item[Keywords:] Instrumental Variables; Nonparametric Identification; Hausman Test.

\item[\emph{JEL classification:}] C26; C14; C21.\newpage

\end{description}
\end{abstract}

\section{Introduction}

The Ordinary Least Squares (OLS) estimator has an appealing nonparametric
interpretation---it provides the optimal linear approximation (in a
mean-square error sense) to the true regression function. That is, the OLS
estimand is a meaningful and easily interpretable parameter under
misspecification of the linear model. Unfortunately, except in special
circumstances (such as with random assignment), this parameter does not have a
causal interpretation. Commonly used estimands based on Instrumental Variables
(IV) do have a causal interpretation (see, e.g., \cite{LATE}), but they do not
share with OLS the appealing nonparametric interpretation (see, e.g.,
\cite{AIG_2000}). The main goal of our paper is to fill this gap and to
propose an IV estimand that has the same nonparametric interpretation as OLS,
but under endogeneity.

The parameter of interest is thus the vector of slopes in the optimal linear
approximation of the \textit{structural} regression function. We call this
parameter the Optimal Linear IV Approximation (OLIVA). We investigate regular
identification of the OLIVA, i.e. identification with a finite efficiency
bound, based on the results in \cite{Severini_Tripathi_2012}. The main
contribution of our paper is to show that a necessary condition for regular
identification of the OLIVA is also sufficient for existence of an IV estimand
in a linear structural regression. That is, we show that, under a minimal
condition for standard inference on the OLIVA, it is possible to obtain an IV
estimator for it.

The identification result is constructive and leads to a Two-Step IV (TSIV)
estimation strategy. The necessary condition for regular identification is a
conditional moment restriction that is used to estimate a suitable instrument
in a first step. The second step is simply a standard linear IV estimator with
the estimated instrument from the first step. The situation is somewhat
analogous to optimal IV (see, e.g., \cite{Robinson} and \cite{Newey_1990}),
but more difficult due to the lack of identification of the first step and the
first step problem being statistically harder than a nonparametric regression
problem. To select an instrument among potentially many candidates, we use
Tikhonov regularization, combined with a sieve approach to obtain a Penalized
Sieve Minimum Distance (PSMD) first step estimator (cf. \cite{Chen_Pouzo_2012}%
). The instrument choice based on Tikhonov is statistically and empirically
justified. Statistically, a Tikhonov instrument exhibits a certain sufficiency
property explained below. Empirically, the resulting PSMD estimator can be
computed with standard regression routines. The TSIV estimator is shown to be
asymptotically normal and to perform favorably in simulations when compared to
alternative estimators, being competitive with the oracle IV under linearity
of the structural model, while robustifying it otherwise.

An important application of our approach is to a Hausman test for exogeneity
that is robust to misspecification of the linear model. This robustness comes
from our TSIV being nonparametrically comparable to OLS under exogeneity. The
robust Hausman test is a standard t-test in an augmented regression that does
not require any correction for standard errors for its validity, as we show
below. \cite{Lochner_Moretti_2015} consider a different exogeneity test
comparing the classical IV estimator with a weighted OLS estimator when the
endogenous variable is discrete. In contrast, our test compares the standard
OLS with our TSIV estimator--more in the spirit of the original \cite{Hausman}%
's exogeneity test--while allowing for general endogenous variables
(continuous, discrete or mixed). Monte Carlo simulations confirm the
robustness of the proposed Hausman test, and the inability of the standard
Hausman test to control the empirical size under misspecification of the
linear model.

Our paper contributes to two different strands of the literature. The first
strand is the nonparametric IV literature; see, e.g., \cite{Newey_Powell_2003}%
, \cite{Ai_Chen_2003}, \cite{HH_2005}, \cite{BCK_2007}, \cite{H_2007},
\cite{H_2011}, \cite{DFFR_2011}, \cite{Santos_2012},
\cite{Chetverikov_Wilhem_2017}, and \cite{Freyberger}, among others.
\cite{Severini_Tripathi_2006, Severini_Tripathi_2012} discuss identification
and efficiency of linear functionals of the structural function without
assuming completeness. Their results on regular identification are adapted to
the OLIVA below. \cite{Santos_2011} establishes regular asymptotic normality
for weighted integrals of the structural function in nonparametric IV, also
allowing for lack of nonparametric identification of the structural function.
\cite{BH_2007} develop a nonparametric test of exogeneity under the maintained
assumption of nonparametric identification$.$ The OLIVA functional was not
considered in \cite{Severini_Tripathi_2006, Severini_Tripathi_2012} or
\cite{Santos_2011}, and the semiparametric robust Hausman test complements the
nonparametric test of \cite{BH_2007}.

Our paper is also related to the Causal IV literature that interprets IV
nonparametrically as a Local Average Treatment Effect (LATE); see \cite{LATE}.
A forerunner of our paper is \cite{Abadie_2003}. He defines the Complier
Causal Response Function and its best linear approximation in the presence of
covariates. He also develops two-step inference for the linear approximation
coefficients when the endogenous variable is binary. Within this binary case,
we show that the OLIVA's slope parameter is the IV estimand resulting from
using the propensity score as instrument, a recommended IV estimator in the
literature (see \cite{LATE} and pg. 623 in \cite{Wooldridge2}). Our asymptotic
results for the binary endogenous case can thus be viewed as extensions of
existing methods (such as, e.g., Theorem 3 in \cite{LATE}) to a
nonparametrically estimated propensity score.

The main theoretical contributions of this paper are thus the interpretation
of the regular identification of the OLIVA as existence of an IV estimand, the
asymptotic normality of a TSIV estimator, and the robust Hausman test. The
identification, estimation and exogeneity test of this paper are all robust to
the lack of the identification of the structural function (i.e. lack of
completeness) and lack of identification of the first step instrument.
Furthermore, the proposed methods are also robust to misspecification of
linear model, sharing the nonparametric robustness of OLS, but in a setting
with endogenous regressors.

We illustrate the utility of our methods with an empirical application
estimating the elasticity of intertemporal substitution (EIS) with quarterly
US data, revisiting previous work by \cite{Yogo}. If the structural
relationship between consumption growth and interest rates is linear, then the
TSIV and standard IV estimands should be the same. In contrast, we obtain a
TSIV estimate much larger than the standard IV estimate, with a similar level
of precision, thereby suggesting that nonlinearities matter in this
application. The TSIV and OLS estimates are rather close, and the robust
Hausman test fails to reject the null hypothesis of exogeneity of real
interest rates.

The rest of the paper is organized as follows. Section \ref{OLIVA} defines
formally the parameter of interest and its regular identification. Section
\ref{TSIV_Estimation} proposes a PSMD first step and establishes the
asymptotic normality of the TSIV. Section \ref{HausmanTest} derives the
asymptotic properties of the robust Hausman test for exogeneity. The finite
sample performance of the TSIV and the robust Hausman test is investigated in
Section \ref{MC}. Finally, Section \ref{EIS} reports the results of our
empirical application to the EIS.\linebreak Appendix A presents notation,
assumptions and some preliminary results that are needed for the main proofs
in Appendix B. A Supplemental Appendix contains further simulation results,
including extensive sensitivity analysis.

\section{Optimal Linear Instrumental Variables Approximations}

\label{OLIVA}

\subsection{Nonparametric Interpretation}

Let the dependent variable $Y$ be related to the $p-$dimensional vector $X$
through the equation%
\begin{equation}
Y=g(X)+\varepsilon, \label{1}%
\end{equation}
where $E[\left.  \varepsilon\right\vert Z]=0$ almost surely (a.s), for a
$q-$dimensional vector of instruments $Z.$

The OLIVA parameter $\beta$ solves, for $g$ satisfying (\ref{1}),
\begin{equation}
\beta=\arg\min_{\gamma\in\mathbb{R}^{p}}E[\left(  g(X)-\gamma^{\prime
}X\right)  ^{2}], \label{BLA}%
\end{equation}
where henceforth $A^{\prime}$ denotes the transpose of $A$. Note that $X$ may
(and in general, will) contain an intercept. For extensions to weighted least
squares versions of (\ref{BLA}) see Section \ref{WLS}.

If $E[XX^{\prime}]$ is positive definite, then%
\begin{equation}
\beta\equiv\beta(g)=E[XX^{\prime}]^{-1}E[Xg(X)]. \label{BLA2}%
\end{equation}
When $X$ is exogenous, i.e. $E[\left.  \varepsilon\right\vert X]=0$ a.s., the
function $g(\cdot)$ is the regression function $E[\left.  Y\right\vert
X=\cdot]$ and $\beta$ is identified and consistently estimated by OLS under
mild conditions. In many economic applications, however, $X$ is endogenous,
i.e. $E[\left.  \varepsilon\right\vert X]\neq0$, and identification and
estimation of (\ref{BLA2}) becomes a more difficult issue than in the
exogenous case, albeit less difficult than identification and estimation of
the structural function $g$ in (\ref{1}). Of course, if $g$ is linear
$g(x)=\gamma_{0}^{\prime}x,$ then $\beta=\gamma_{0}.$

We first investigate regular identification of $\beta$ in (\ref{1}%
)-(\ref{BLA}). The terminology of regular identification is proposed in
\cite{KhanTamer}, and refers to identification with a finite efficiency bound.
Regular identification of a parameter is desirable because it means
possibility of standard inference (see \cite{Chamberlain86}). It will be shown
below that a necessary condition for regular identification of $\beta$ is%
\begin{equation}
E[\left.  h(Z)\right\vert X]=X\text{ a.s,} \label{Reg}%
\end{equation}
for an square integrable $h(\cdot);$ see Lemma \ref{Lemma1}, which builds on
\cite{Severini_Tripathi_2012}. We show that condition (\ref{Reg}) is also
sufficient for existence of an IV estimand identifying $\beta.$ That is, we
show that (\ref{Reg}) implies that $\beta$ is identified from a linear
structural regression%
\begin{equation}
Y=X^{\prime}\beta+U,\qquad E[Uh(Z)]=0. \label{linearIV}%
\end{equation}
The IV estimand uses the unknown, possibly not unique, transformation
$h(\cdot)$ of $Z$ as instruments. We propose below a Two-Step IV (TSIV)
estimator that first estimates the instruments from (\ref{Reg}) and then
applies IV with the estimated instruments. The proposed IV estimator has the
same nonparametric interpretation as OLS, but under endogeneity.

If the nonparametric structural function $g$ is identified, then $\beta$ is of
course identified (from \ref{BLA2}). Conditions for point identification and
consistent estimation of $g$ are given in the references above on the
nonparametric IV literature. Likewise, asymptotic normality for continuous
functionals of a point-identified $g$ has been analyzed in \cite{Ai_Chen_2003}%
, \cite{Ai_Chen_2007}, \cite{CFR_hb}, \cite{CFR_2014}, \cite{Chen_Pouzo_2015}
and \cite{BreunigJohannes_2016}, among others.

Nonparametric identification of $g$ is, however, not necessary for
identification of the OLIVA; see \cite{Severini_Tripathi_2006,
Severini_Tripathi_2012}. It is indeed desirable to obtain identification of
$\beta$ without requiring completeness assumptions, which are known to be
impossible to test (cf. \cite{Canay_Santos_Shaikh}). In this paper we focus on
regular identification of the OLIVA without assuming completeness, i.e.
without assuming identification of $g$.

Section \ref{RegIden} below shows the necessity of the conditional moment
restriction (\ref{Reg}) for regular identification of the OLIVA. When regular
identification of the OLIVA does not hold, but the OLIVA is identified, we
expect our estimator to provide a good approximation to the OLIVA. This
follows because (i) under irregular identification of the OLIVA, the first
step instrument approximately solves the first step conditional moment, and
(ii) small errors in the first step equation lead to small errors in the
second step limit.\footnote{We thank Andres Santos for making this point to
us.} Inference under irregular identification is known to be less stable, see
\cite{Chamberlain86}, and it is beyond the scope of this paper. See
\cite{Babii_Florens} for recent advances in this direction, and
\cite{Escanciano_Li_Bounds} for partial identification results.

\subsection{Regular Identification of the OLIVA}

\label{RegIden}

We observe a random vector $W=(Y,X^{\prime},Z^{\prime})^{\prime}$ satisfying
(\ref{1}), or equivalently,
\begin{equation}
r(z):=E[\left.  Y\right\vert Z=z]=E[\left.  g(X)\right\vert Z=z]:=T^{\ast
}g(z), \label{2}%
\end{equation}
where $T^{\ast}$ denotes the adjoint operator of the operator $T,$ with
$Th(x)=E[\left.  h(Z)\right\vert X=x]$ a.s. Let $\mathcal{G}$ denote the
parameter space for $g.$ Assume $g\in\mathcal{G}\subseteq L_{2}(X)$ and $r\in
L_{2}(Z),$ where henceforth, for a generic random variable $V,$ $L_{2}(V)$
denotes the space of (measurable) square integrable functions of $V,$ i.e.
$f\in L_{2}(V)$ if $\left\Vert f\right\Vert ^{2}:=E\left[  \left\vert
f(V)\right\vert ^{2}\right]  <\infty,$ and where $\left\vert A\right\vert
=trace\left(  A^{\prime}A\right)  ^{1/2}$ is the Euclidean norm.\footnote{When
$f$ is vector-valued, by $f(V)\in L_{2}(V)$ we mean that its components are
all in $L_{2}(V).$}

The next result, which follows from an application of Lemma 4.1 in
\cite{Severini_Tripathi_2012}, provides a necessary condition for regular
identification of the OLIVA. Define $g_{0}:=\arg\min_{g:r=T^{\ast}g}\left\Vert
g\right\Vert ,$ and note that correct specification of the model guarantees
that $g_{0}$ is uniquely defined; see \cite{EHN}. Define $\xi=Y-g_{0}(X)$,
$\Omega(z)=E[\left.  \xi^{2}\right\vert Z=z],$ and let $\mathcal{S}_{Z}$
denote the support of $Z.$ For future reference, define the range of the
operator $T$ as $\mathcal{R}(T):=\{f\in L_{2}(X):\exists s\in L_{2}%
(Z),Ts=f\},$ and for a subspace $V,$ let $V^{\perp}$ and $\overline{V}$
denote, respectively, its orthogonal complement and its closure. \bigskip

\noindent\textbf{Assumption 1}: (\ref{2}) holds, $g\in\mathcal{G}\subseteq
L_{2}(X),$ $r\in L_{2}(Z),$ and $E[XX^{\prime}]$ is finite and positive
definite$.$\bigskip

\noindent\textbf{Assumption 2}: $0<\inf_{z\in\mathcal{S}_{Z}}\Omega(z)\leq
\sup_{z\in\mathcal{S}_{Z}}\Omega(z)<\infty$ and $T$ is compact.\bigskip

\noindent\textbf{Assumption 3}: There exists $h(\cdot)\in L_{2}(Z)$ such that
(\ref{Reg}) holds.

\begin{lemma}
\label{Lemma1}Let Assumptions 1-2 hold. If $\beta$ is regularly identified,
then Assumption 3 must hold.
\end{lemma}

\noindent The proof of Lemma \ref{Lemma1} and other results in the text are
gathered in Appendix B. Assumptions 1 and 2 are taken from
\cite{Severini_Tripathi_2012} and are standard in the literature. Given the
necessity of Assumption 3 and its importance for our results it is useful to
provide some discussion on it. The first observation is that although
sufficient conditions for Assumption 3 to hold can be obtained for parametric
settings, such as those in the Monte Carlo section, it is hard to give
primitive sufficient conditions in nonparametric settings. The second
observation is that Assumption 3 may hold when $L_{2}-$completeness of $X$
given $Z$ fails and $g$ is thus not identified (see \cite{Newey_Powell_2003}
for discussion of $L_{2}-$completeness). To illustrate this point, we consider
the empirically relevant case where $X$ is continuous and $Z$ is discrete. It
is well known that in this case $g$ is not identified. In contrast, Assumption
3 may hold and, importantly, it is testable. To see this, let $\{z_{1}%
,...,z_{J}\}$ denote the support of $Z,$ with $J<\infty,$ and note that any
function $h$ can be identified with a $J\times p$ matrix through the
representation%
\[
h(z)=\sum_{j=1}^{J}h(z_{j})1(z=z_{j}),
\]
where $1(A)$ is the indicator function of the event $A.$ Assumption 3 is then
simply the conditional moment restriction with a finite number of parameters
$\theta=(h(z_{1}),...,h(z_{J}))\in\mathbb{R}^{p\times J}$ given by%
\begin{equation}
E[\left.  \theta\mathbf{1}-X\right\vert X]=0\text{ a.s.} \label{CMR}%
\end{equation}
where $\mathbf{1}=(1(Z=z_{1}),...,1(Z=z_{J}))^{\prime}.$ To deal with the
potential lack of identification of $h$ (i.e. of $\theta)$ we use the minimum
norm estimator described below, which is consistent for a population analog
$h_{0}(Z)=\theta_{0}\mathbf{1}$. Furthermore, the estimator of $\theta_{0}$
can be shown to be asymptotically normal. Thus, standard tools from
nonparametric regression testing can be used to test for (\ref{CMR}); see,
e.g., \cite{Bierens} and \cite{Escanciano_2006}. Whether the nonparametric
conditional moment restriction in Assumption 3 is testable more generally
(i.e. with continuous $Z)$ is a delicate issue, see \cite{ChenSantos}, and it
will be investigated elsewhere.

When Assumption 3 does not hold two possibilities may arise: (i) $\beta$ is
identified, but it has infinite efficiency bound, and (ii) $\beta$ is not
identified. When $\beta$ is identified and Assumption 3 fails, $X$ belongs to
the boundary of the range of $T$ (i.e. $X\in\overline{\mathcal{R}(T)}%
\setminus\mathcal{R}(T),$ see \cite{Severini_Tripathi_2012}), and thus our IV
estimand can be made arbitrarily close to $\beta$. As we explain below in
Remark \ref{A3Robust}, even when Assumption 3 does not hold, our estimator has
a well-defined IV estimand as its limit, provided a mild condition is satisfied.

The main observation of this paper is that the necessary condition for regular
identification of $\beta$ is also sufficient for existence of an IV estimand.
This follows because by the law of iterated expectations, Assumption 3 and
$E[\left.  \varepsilon\right\vert Z]=0$ a.s.,
\begin{align}
\beta &  =E[XX^{\prime}]^{-1}E[Xg(X)]\nonumber\\
&  =E[E[\left.  h(Z)\right\vert X]X^{\prime}]^{-1}E[E[\left.  h(Z)\right\vert
X]g(X)]\nonumber\\
&  =E[h(Z)X^{\prime}]^{-1}E[h(Z)Y], \label{IVest}%
\end{align}
which is the IV estimand using $h(Z)$ as instruments for $X.$ We note that to
obtain this IV representation in (\ref{IVest}) a weaker exogeneity than
$E[\left.  \varepsilon\right\vert Z]=0$ suffices, namely $E[\varepsilon
h(Z)]=0.$ We maintain the \textquotedblleft strict\textquotedblright%
\ exogeneity $E[\left.  \varepsilon\right\vert Z]=0$ because it is often used
in the literature and simplifies some of our subsequent asymptotic results,
although see Remark \ref{weakexo}. The following result summarizes this
finding and shows that, although there are potentially many solutions to
(\ref{Reg}), the corresponding $\beta$ is unique.

\begin{proposition}
\label{Proposition 1}Let Assumptions 1-3 hold. Then, $\beta$ is regularly
identified as (\ref{IVest})$.$
\end{proposition}

\begin{remark}
By (\ref{Reg}), $E[h(Z)X^{\prime}]=E[XX^{\prime}].$ Thus, non-singularity of
$E[h(Z)X^{\prime}]$ follows from that of $E[XX^{\prime}].$ Thus, the strength
of the instruments $h(Z)$ is measured by the level of multicollinearity in $X$.
\end{remark}

\section{Two-Step Instrumental Variables Estimation}

\label{TSIV_Estimation}

Proposition \ref{Proposition 1} suggests a TSIV estimation method where,
first, an $h$ is estimated from (\ref{Reg}) and then, an IV estimator is
considered using the estimated $h$ as instrument. To describe the estimator,
let $\{W_{i}\equiv(Y_{i},X_{i}^{\prime},Z_{i}^{\prime})^{\prime}\}_{i=1}^{n}$
be an independent and identically distributed $(iid)$ sample of size $n$
satisfying (\ref{1}). The TSIV estimator follows the steps:

\begin{description}
\item[Step 1.] Estimate an instrument $h(Z)$ satisfying $E[\left.
h(Z)\right\vert X]=X$ a.s., say $\hat{h}_{n},$ as defined in (\ref{sol}) below.

\item[Step 2.] Run linear IV using instruments $\hat{h}_{n}(Z)$ for $X$ in
$Y=X^{\prime}\beta+U,$ i.e.
\begin{equation}
\hat{\beta}=\left(  \frac{1}{n}\sum_{i=1}^{n}\hat{h}_{n}(Z_{i})X_{i}^{\prime
}\right)  ^{-1}\left(  \frac{1}{n}\sum_{i=1}^{n}\hat{h}_{n}(Z_{i}%
)Y_{i}\right)  , \label{TSIV}%
\end{equation}
where $\hat{h}_{n}$ is the first step estimator given in Step 1.
\end{description}

For ease of exposition, we consider first the case where $X$ and $Z$ have no
overlapping components (i.e. no included exogenous or controls) and both are
continuous. We also analyze below the case of control variables and discrete variables.

\subsection{First-Step Estimation}

As argued in pg. 130 of \cite{Santos_2012} identification of $h$ in
(\ref{Reg}) is problematic, as in most instances instruments posses a
variation that is unrelated to the endogenous regressor (i.e., there exists a
function $\psi(z)$ such that $E[\left.  \psi(Z)\right\vert X]=0$ a.s.). To
deal with the problem of lack of uniqueness of $h,$ we consider a
Tikhonov-type estimator. This approach is commonly used in the literature
estimating $g$, see \cite{HH_2005}, \cite{CFR_hb}, \cite{FJV_2011},
\cite{Chen_Pouzo_2012} and \cite{Gagliardini_Scaillet_2012}, among others.
\cite{Chen_Pouzo_2012} propose a PSMD estimator of $g$ and show the $L_{2}%
-$consistency of a solution identified via a strict convex penalty. These
authors also obtain rates in Banach norms under point identification. Our
first-step estimator $\hat{h}_{n}$ is a PSMD estimator of the form considered
in \cite{Chen_Pouzo_2012} when identification is achieved with an $L_{2}%
$-penalty. As it turns out, the Tikhonov-type or $L_{2}$-penalty estimator is
well motivated in our setting, as we explain below. It implies that our
instrument satisfies a certain sufficiency property.

Defining $m(X;h):=E[h(Z)-X|X]$, we estimate the unique $h_{0}$ satisfying
$h_{0}=\lim_{\lambda\downarrow0}h_{0}(\lambda),$ where%
\[
h_{0}(\lambda)=\arg\min\{||m(\cdot;h)||^{2}+\lambda||h||^{2}:h\in L_{2}(Z)\},
\]
and $\lambda>0.$ Assumption 3 guarantees the existence and uniqueness of
$h_{0},$ see \cite{EHN}. The sufficiency property mentioned above is that for
any distinct solution $h_{1}$ of (\ref{Reg}), $h_{1}\neq h_{0},$ it holds that
in the first stage regression%
\begin{equation}
X=c_{0}+\alpha_{0}h_{0}(Z)+\alpha_{1}h_{1}(Z)+V,\qquad Cov(V,h_{j}%
(Z))=0,\text{ }j=0,1, \label{first-stage}%
\end{equation}
$\alpha_{1}$ must be zero, as shown in the next result. We note that $V$ is
simply a least squares (i.e. reduced form) error term. \bigskip

\begin{proposition}
\label{Proposition 3}Let $h_{1}\neq h_{0}$ be another solution of (\ref{Reg}).
Then, $\alpha_{1}=0$ in (\ref{first-stage}).
\end{proposition}

This result states that after controlling for $h_{0}(Z)$ in the first stage
regression, any other distinct solution to (\ref{Reg}) is irrelevant in the
first stage. It is in this precise sense that we say $h_{0}(Z)$ is sufficient.
We note, however, that this property does not imply that $h_{0}$ is better
than any other solution to (\ref{Reg}) in terms of leading to a more efficient
estimation of $\beta$. For efficiency considerations see
\cite{Severini_Tripathi_2012}.

\begin{remark}
\label{A3Robust}The minimum norm $h_{0}$ is well-defined under a weaker
condition than Assumption 3. From \cite{EHN}, for existence of $h_{0}$ it
suffices that $X$ belongs to the dense set $\mathcal{R}(T)+\mathcal{R}%
(T)^{\bot}.$ In particular, this assumption holds when $X$ is a square
integrable continuous variable and $Z$ is discrete (since $\mathcal{R}%
(T)+\mathcal{R}(T)^{\bot}\equiv L_{2}(X)$ in this case). Thus, under mild
conditions, $\hat{\beta}$ has a probabilistic limit satisfying (\ref{linearIV}).
\end{remark}

Having motivated the Tikhonov-type instrument, we introduce now its PSMD
estimator. Let $E_{n}[g(W)]$ denote the sample mean operator, i.e.
$E_{n}[g(W)]=n^{-1}\sum_{i}^{n}g(W_{i}),$ let $||g||_{n}=\left(
E_{n}[\left\vert g(W)\right\vert ^{2}]\right)  ^{1/2}$ be the empirical
$L_{2}$ norm, and let $\hat{E}[\left.  h(Z)\right\vert X]$ be a series-based
estimator for the conditional mean $E[\left.  h(Z)\right\vert X],$ which is
given as follows. Consider a vector of approximating functions%
\[
p^{K_{n}}(x)=(p_{1}(x),...,p_{K_{n}}(x))^{\prime},
\]
having the property that a linear combination can approximate $E[\left.
h(Z)\right\vert X=x]$ well. Then,
\[
\hat{E}[\left.  h(Z)\right\vert X=x]={p^{K_{n}}}^{\prime}(x)(P^{\prime}%
P)^{-1}\sum_{i=1}^{n}p^{K_{n}}(X_{i})h(Z_{i}),
\]
where $P=[p^{K_{n}}(X_{1}),...,p^{K_{n}}(X_{n})]^{\prime}$ and $K_{n}%
\rightarrow\infty$ as $n\rightarrow\infty$.

Let $\mathcal{H}\subseteq L_{2}(Z)$ denote the parameter space for $h.$ Then,
define the estimator%
\begin{equation}
\hat{h}_{n}:=\arg\min\{||\hat{m}(X;h)||_{n}^{2}+\lambda_{n}||h||_{n}^{2}%
:h\in\mathcal{H}_{n}\}, \label{Tik}%
\end{equation}
where $\mathcal{H}_{n}\subset\mathcal{H}\subseteq L_{2}(Z)$ is a linear sieve
parameter space whose complexity grows with sample size$,$ $\hat{m}%
(X_{i};h)=\hat{E}(h(Z)-X|X_{i})$, and $\lambda_{n}$ is a sequence of positive
numbers satisfying that $\lambda_{n}\downarrow0$ as $n\uparrow\infty,$ and
some further conditions given in the Appendix A. In our implementation
$\mathcal{H}_{n}$ is the finite dimensional linear sieve given by
\begin{equation}
\mathcal{H}_{n}=\left\{  h:h=\sum_{j=1}^{J_{n}}a_{j}q_{j}(\cdot)\right\}
\label{Hn}%
\end{equation}
where $q^{J_{n}}(z)=(q_{1}(z),...,q_{J_{n}}(z))^{\prime}$ is a vector
containing a linear sieve basis, with $J_{n}\rightarrow\infty$ as
$n\rightarrow\infty$.

To better understand the first step estimator and how it can be computed by
standard methods consider the approximation%
\[
X=E[\left.  h(Z)\right\vert X]\approx E[\left.  a^{\prime}q^{J_{n}%
}(Z)\right\vert X]=a^{\prime}E[\left.  q^{J_{n}}(Z)\right\vert X],
\]
which suggests a two step procedure for obtaining $\hat{h}_{n}:$ (i) first
compute the fitted values $\hat{q}(X)=\hat{E}[\left.  q^{J_{n}}(Z)\right\vert
X]$ by OLS of $q^{J_{n}}(Z)$ on $p^{K_{n}}(X);$ and then (ii) run Ridge
regression $X$ on $\hat{q}(X).$ Indeed, if we define $D_{n}=E_{n}[\hat
{q}(X)X^{\prime}]$, $Q_{2n}=E_{n}[q^{J_{n}}(Z)q^{J_{n}}(Z)^{\prime}],$ and%
\[
\hat{A}_{\lambda_{n}}=E_{n}[\hat{q}(X)\hat{q}(X)^{\prime}]+\lambda_{n}Q_{2n}.
\]
Then, the closed form solution to (\ref{Tik}) is given by%
\begin{equation}
\hat{h}_{n}(\cdot)=D_{n}^{\prime}\hat{A}_{\lambda_{n}}^{-1}q^{J_{n}}(\cdot).
\label{sol}%
\end{equation}
This estimator can be easily implemented by an OLS and a \emph{standard} Ridge
regression steps: (i) standardize $q^{J_{n}}$ so that $Q_{2n}$ becomes the
identity (simply multiply the original $q^{J_{n}}$ by $Q_{2n}^{-1/2});$ (ii)
run OLS $q^{J_{n}}(Z)$ on $p^{K_{n}}(X)$ and keep fitted values $\hat{q}(X);$
(iii) run standard Ridge regression of $X$ on $\hat{q}(X);$ the slope
coefficient in the last regression is $D_{n}^{\prime}\hat{A}_{\lambda_{n}%
}^{-1}.$ Section \ref{Implementation} further discusses implementation of the
estimation of $h_{0}$ in a more general setting with additional exogenous variables.

An alternative minimum norm approach requires choosing two sequences of
positive numbers $a_{n}$ and $b_{n}$ and solving the program
\[
\tilde{h}_{n}:=\arg\min\{||h||_{n}^{2}:h\in\mathcal{H}_{n},||\hat
{m}(X;h)||_{n}^{2}\leq b_{n}/a_{n}\}.
\]
This is the approach used in \cite{Santos_2011} for different functionals than
the OLIVA. We prefer our implementation, since we only need one tuning
parameter rather than two, and data driven methods for choosing $\lambda_{n}$
are readily available; see Section \ref{Implementation}.

\subsection{Second-Step Estimation and Inference}

This section establishes the consistency and asymptotic normality of
$\hat{\beta},$ and the consistency of its asymptotic variance, which is useful
for inference. Recall $W=(Y,X^{\prime},Z^{\prime})^{\prime}$ and define%
\begin{equation}
m(W,\beta,h,g)=(Y-X^{\prime}\beta)h(Z)-(g(X)-X^{\prime}\beta)(h(Z)-X),
\label{inf}%
\end{equation}
with the short notation $m_{0}=m(W,\beta,h_{0},g_{0}).$ The second term in
(\ref{inf}) accounts for the asymptotic impact of estimating the instrument
$h_{0}.$ When the minimum norm structural function $g_{0}$ is linear, like
with a binary treatment, this second term is zero and there will be no impact
from estimating $h_{0}$ on inference. Thus, we can interpret this second term
in $m$ as accounting for a \textquotedblleft nonlinearity
bias\textquotedblright\ in inference of the IV estimator.

To estimate the asymptotic variance of $\hat{\beta}$ is useful to estimate
$g_{0}$. We introduce a Tikhonov-type estimator that is the dual of $\hat
{h}_{n}$. Let $\hat{g}_{n}(\cdot)$ denote a PSMD estimator of $g_{0}$ given by%
\begin{equation}
\hat{g}_{n}(\cdot)=G_{n}^{\prime}B_{\lambda_{n}}^{-1}p^{K_{n}}(\cdot),
\label{ghat}%
\end{equation}
with $G_{n}=E_{n}[\hat{p}(Z)Y],$ $\hat{p}(Z)=\hat{E}[\left.  p^{K_{n}%
}(X)\right\vert Z],$ $\hat{E}[\left.  g(X)\right\vert Z=z]={q^{J_{n}}}%
^{\prime}(z)(Q^{\prime}Q)^{-1}\sum_{i=1}^{n}q^{J_{n}}(Z_{i})g(X_{i})$,
$Q=[q^{J_{n}}(Z_{1}),...,q^{J_{n}}(Z_{n})]^{\prime},$ $P_{2n}=E_{n}[p^{K_{n}%
}(X)p^{K_{n}}(X)^{\prime}],$ and $\hat{B}_{\lambda_{n}}=E_{n}[\hat{p}%
(Z)\hat{p}(Z)^{\prime}]+\lambda_{n}P_{2n}.$ For ease of presentation, we use
the same notation for the tuning parameters in $\hat{h}_{n}$ and $\hat{g}%
_{n},$ although of course we will use different tuning parameters $K_{n}$ and
$J_{n}$ for estimating $\hat{h}_{n}$ or $\hat{g}_{n},$ see Section
\ref{Implementation} for issues of implementation.

\begin{theorem}
\label{ANTSIV}Let Assumptions 1-3 above and Assumptions A1-A5, A6(i-iii) in
the Appendix \textit{A} hold. Then, $\hat{\beta}$ is consistent and
asymptotically normal, i.e.%
\[
\sqrt{n}(\hat{\beta}-\beta)\longrightarrow_{d}N(0,\Sigma),
\]
where $\Sigma=E[h_{0}(Z)X^{\prime}]^{-1}E[m_{0}m_{0}^{\prime}]E[Xh_{0}%
(Z)^{\prime}]^{-1}$. Furthermore, $\Sigma$ is consistently estimated by
\begin{equation}
\hat{\Sigma}=E_{n}[\hat{h}_{n}(Z_{i})X_{i}^{\prime}]^{-1}E_{n}[\hat{m}%
_{ni}\hat{m}_{ni}^{\prime}]E_{n}[X_{i}\hat{h}_{n}^{\prime}(Z_{i})]^{-1},
\label{Avar}%
\end{equation}
where $\hat{m}_{ni}=m(W_{i},\hat{\beta},\hat{h}_{n},\hat{g}_{n}).$

\begin{remark}
\label{weakexo}If $E[\left.  \varepsilon\right\vert Z]=0$ is relaxed to only
$E[\varepsilon h_{0}(Z)]=0,$ then the asymptotic normality of $\hat{\beta}$
goes through with $(g(X)-X^{\prime}\beta)$ in $m_{0}$ replaced by $v_{n}$ in
(\ref{vn}) of the Appendix, provided $v_{n}$ and the resulting $m_{0}$ have
finite variances, see the proof of Theorem \ref{ANTSIV}.
\end{remark}
\end{theorem}

The assumptions in Theorem \ref{ANTSIV} are standard in the literature of
two-step semiparametric estimators. Theorem \ref{ANTSIV} can be then used to
construct confidence regions for $\beta$ and testing hypotheses about $\beta$
following standard procedures. The proof of Theorem \ref{ANTSIV} relies on new
$L_{2}-$rates of convergence for $\hat{h}_{n}$ and $\hat{g}_{n}$ under partial
identification of $h$ and $g$ (note that the rates in \cite{Chen_Pouzo_2012}
are given under point identification and \cite{Santos_2011} obtained related
rates but for a weak norm).

\subsection{Partial Effects Interpretation, Exogenous Controls and Discrete
Variables}

We provide now a partial effects interpretation for subvectors of the OLIVA
parameter $\beta$ that are analogous to OLS. Define $X=(X_{1}^{\prime}%
,X_{2}^{\prime})^{\prime}$ and partition $\beta$ accordingly as $\beta
=(\beta_{1}^{\prime},\beta_{2}^{\prime})^{\prime}.$ Suppose we are only
interested in $\beta_{2}.$ From standard OLS theory, we obtain
\[
\beta_{2}=E[V_{2}V_{2}^{\prime}]^{-1}E[V_{2}g(X)],
\]
where $V_{2}$ is the OLS error from the regression of $X_{2}$ on $X_{1}.$ This
result could be used to obtain an estimator of $\beta_{2}$ that does not
compute an estimator for $\beta_{1}$ and that reduces the dimensionality of
the problem of estimating $h$ (from the dimension of the original $X$ to the
dimension of $X_{2}$)$,$ since now we can use the condition
\[
E[\left.  h(Z)\right\vert V_{2}]=V_{2}\text{ a.s.}%
\]
This method might be particularly useful when the dimension of $X_{1}$ is
large and $g$ has a partly linear structure%
\begin{equation}
g(X)=\gamma_{1}^{\prime}X_{1}+g_{2}(X_{2}), \label{Partlylinear}%
\end{equation}
since then $\beta_{2}=E[V_{2}V_{2}^{\prime}]^{-1}E[V_{2}g_{2}(X_{2})]$ can be
interpreted as providing a best linear approximation to $g_{2}(X_{2})$ with a
linear function of $V_{2}$, i.e.%
\[
\beta_{2}=\arg\min_{b_{2}}E[\left(  g_{2}(X_{2})-b_{2}^{\prime}V_{2}\right)
^{2}].
\]
In this discussion, $X_{1}$ could be variables that are of secondary interest.

Suppose now that there are exogenous variables included in the structural
equation $g$. This means $X$ and $Z$ have common components. Specifically,
with some abuse of notation, define $X=(X_{1}^{\prime},X_{2}^{\prime}%
)^{\prime}$ and $Z=(Z_{1}^{\prime},Z_{2}^{\prime})^{\prime}$ where
$X_{1}=Z_{1}$ denote the overlapping components of $X$ and $Z,$ with dimension
$p_{1}=q_{1}.$ This is a very common situation in applications, where
exogenous controls are often used. In this setting a solution of $E[\left.
h(Z)\right\vert X]=X$ a.s. has the form $h(Z)=(Z_{1}^{\prime},h_{2}^{\prime
}(Z))^{\prime},$ where
\begin{equation}
E[\left.  h_{2}(Z)\right\vert X]=X_{2}\text{ a.s.} \label{over}%
\end{equation}
Following the arguments of the general case, we could obtain an estimator
given by $\hat{h}_{n}=(Z_{1}^{\prime},\hat{h}_{2n}^{\prime})^{\prime},$ where
\begin{equation}
\hat{h}_{2n}(\cdot)=D_{2n}^{\prime}\hat{A}_{\lambda_{n}}^{-1}q^{J_{n}}(\cdot),
\label{h2}%
\end{equation}
and $D_{2n}:=E_{n}[\hat{q}(X)X_{2}^{\prime}].$ This setting also covers the
case of an intercept with no other common components, where $X_{1}=Z_{1}=1$
and $q_{1}=1.$ The asymptotic normality for $\hat{\beta}$ continues to hold,
with no changes in the asymptotic distribution.

If the dimension of $X_{1}$ is high and the sample size is moderate, the
method above may not perform well due to the curse of dimensionality. Equation
(\ref{over}) implies%
\begin{equation}
E[\left.  h_{2}(Z)\right\vert X_{2}]=X_{2}\text{ a.s.} \label{red}%
\end{equation}
so that nonparametric estimation of $h_{20}$ only involves functions
$p^{K_{n}}(X_{2})$ and $q^{J_{n}}(Z).$ Equation (\ref{red}) is still necessary
for regular identification. Summarizing, for implementing our methods with
moderate or high dimensional controls $X_{1}$ we recommend our general
algorithm above with bases $\{p^{K_{n}}(X_{2})\}$ and $\left\{  q^{J_{n}%
}(Z)\right\}  ,$ which is consistent with the specification in
(\ref{Partlylinear}). Further details on implementation are provided in
Section \ref{Implementation}.

Simplifications occur when some variables are discrete. When the endogenous
variable $X$ is discrete we do not need $K_{n}\rightarrow\infty,$ and we can
choose $p^{K_{n}}$ as a saturated basis. Consider first the important case of
a binary endogenous variable $X=(1,X_{2})$ with $X_{2}\in\left\{  0,1\right\}
$. Define the propensity score $\pi(z):=\Pr\left(  \left.  X_{2}=1\right\vert
Z=z\right)  .$ We show below that under the mild assumption that $\pi(z)$ is
not constant, Assumption 3 holds. Furthermore, the minimum norm solution
$h_{0}$ is simply
\begin{equation}
h_{0}(z)=\alpha+\gamma\pi(z), \label{h0}%
\end{equation}
where $\alpha=\bar{\pi}\left(  1-\gamma\right)  ,$ $\gamma=\bar{\pi}%
(1-\bar{\pi})/var(\pi(Z))$ and $\bar{\pi}=\Pr\left(  X_{2}=1\right)  $. An
implication of this representation is that the slope of the OLIVA is
\begin{equation}
\frac{Cov(Y,h_{0}(Z))}{Cov(X_{2},h_{0}(Z))}=\frac{Cov(Y,\pi(Z))}{Cov(X_{2}%
,\pi(Z))}\equiv\alpha_{\pi}^{IV}, \label{slope}%
\end{equation}
i.e., the LATE estimand $\alpha_{\pi}^{IV}$ using the propensity score as
instrument, which was suggested in \cite{LATE}. Thus, the OLIVA\ in the binary
endogenous case coincides with an important IV estimand recommended in the
literature. We summarize our findings in the following result. The proof can
be found in the Appendix.

\begin{proposition}
\label{Proposition 2}If $X=(1,X_{2})$ with $X_{2}$ a binary endogenous
variable, $0<\bar{\pi}<1,$ and $var(\pi(Z))>0,$ then Assumption 3 holds with a
minimum norm solution $h_{0}$ given by (\ref{h0}). Furthermore, the OLIVA is
$\beta=(c_{\pi}^{IV},\alpha_{\pi}^{IV})^{\prime},$ where $c_{\pi}%
^{IV}=E[Y]-\alpha_{\pi}^{IV}\bar{\pi}$ and $\alpha_{\pi}^{IV}$ is defined in
(\ref{slope}).
\end{proposition}

This result implies that for the binary endogenous case estimating $h_{0},$
and then $\beta_{0},$ simply requires estimating nonparametrically the
propensity score.

More generally, if $X$ has $d$ points of support, say $\left\{  x_{1}%
,...,x_{d}\right\}  ,$ then we can set $K_{n}=d$ and $p_{k}(x)=1(x=x_{k}),$
$k=1,..,K_{n},$ in our general algorithm. Define the unconditional
probabilities $\Pr\left(  X=x_{j}\right)  =\pi_{j},$ $j=1,...,d.$ Then,
Assumption 3 boils down to existence of $h$ satisfying the linear equalities,
for $k=1,...,d,$%
\begin{equation}
E[h(Z)p_{k}(X)]=\pi_{k}x_{k}. \label{disc}%
\end{equation}
Using Theorem 2, pg. 65, in \cite{Luenberger_Book}, we can find a closed form
solution for $h_{0}$ as follows. Define the generalized propensity scores
$\pi_{j}(z):=\Pr\left(  \left.  X=x_{j}\right\vert Z=z\right)  $ and the
random vector $\Pi\equiv\Pi(Z)=(\pi_{1}(Z),...,\pi_{d}(Z))^{\prime}.\ $If
$E[\Pi\Pi^{\prime}]$ is positive definite, then the minimum norm solution to
(\ref{Reg}) is given by $h_{0}(z)=\gamma^{\prime}\Pi(z)$ where $\gamma=\left(
E[\Pi\Pi^{\prime}]\right)  ^{-1}S$ and $S=(\pi_{1}x_{1},...,\pi_{d}%
x_{d})^{\prime}.$ Thus, for discrete endogenous variables our nonparametric
algorithm with $K_{n}=d$ and $p_{k}(x)=1(x=x_{k}),$ $k=1,..,K_{n},$ is a
semiparametric method where $h_{0}(z)=\gamma^{\prime}\Pi(z)$ is estimated by
estimating the conditional probabilities $\Pi(z)$ by $\hat{\Pi}(z)=(\hat{\pi
}_{1}(Z),...,\hat{\pi}_{d}(Z)),$ where $\hat{\pi}_{k}(z)=\hat{E}[\left.
p_{k}(X)\right\vert Z=z].$ In estimating $\gamma,$ if the sample analog of
$E[\Pi\Pi^{\prime}]$ is positive definite, then there is no need to choose
$\lambda$ for estimating $h_{0}.$ If this matrix is not invertible, we can
apply the Tikhonov-type estimator, as proposed above.

Similarly, when $Z$ is discrete we do not need $J_{n}$ diverging to infinity.
As before, we can choose a linear sieve $\mathcal{H}_{n}$ that is saturated
and $q^{J_{n}}(Z)$ could be a saturated basis for it. Specifically, if $Z$
takes $J$ discrete values, $\{z_{1},...,z_{J}\}$, we can take $q_{j}%
(z)=1(z=z_{j}),$ $j=1,...,J_{n}\equiv J.$

Summarizing, all the different cases (with or without controls, nonparametric
or semiparametric structural functions, discrete or continuous variables) can
be implemented with the same algorithm but with different definitions of the
approximation bases $\{p^{K_{n}}(X),q^{J_{n}}(Z)\}.$ In all these cases, the
formulas for the asymptotic variance of $\hat{\beta}$ remain the same. The
following section provides further details on implementation.

\subsection{Implementation}

\label{Implementation}

To enhance the practical applicability of our method we summarize its
implementation in what we think is the most useful case in empirical
applications: estimation in the presence of a vector of controls entering
linearly in the models for $g$ and $h$. Since the vector of controls can be
high dimensional, we do not think of the linearity of the controls as a strong
assumption. As before, we split $X=(X_{1}^{\prime},X_{2}^{\prime})^{\prime}$
and $Z=(Z_{1}^{\prime},Z_{2}^{\prime})^{\prime},$ where $X_{1}=Z_{1}$ denote
the vector of exogeneous controls (containing an intercept), with dimension
$p_{1}=q_{1}.$ The endogenous variable of interest $X_{2}$ has dimension
$p_{2}=p-p_{1},$ and the instrument $Z_{2}$ has dimension $q_{2}=q-q_{1}.$
Following the discussion above, for implementation one has to choose bases
$\{p^{K_{n}}(x),q^{J_{n}}(z)\}$ and the tuning parameters $\{J_{n}%
,K_{n},\lambda_{n}\}.$ Using these imputs, we estimate an instrument $\hat
{h}_{n}=(Z_{1}^{\prime},\hat{h}_{2n}^{\prime})^{\prime},$ where $\hat{h}_{2n}$
estimates a minimum norm solution $h_{20}$ of
\[
E[\left.  h_{2}(Z)\right\vert X_{2}]=X_{2}\text{ a.s.}%
\]
An appealing feature of sieve estimation is that additional semiparametric
restrictions can be imposed on $h_{2}$ simply by restricting the terms in the
basis $\{q^{J_{n}}(z)\}.$ These include additivity or exclusion restrictions,
among others. For example, one restriction that we impose in this section is
that $h_{2}$ is linear in $Z_{1},$ i.e. $h_{2}(Z)=a_{1}^{\prime}Z_{1}%
+h_{20}(Z_{2}).$ This is, of course, not necessary for regular identification,
but it ameliorates the curse of dimensionality, specially when $Z_{1}$ is high
dimensional, and it may lead to better finite sample performance (by reducing variance).

As explained in the previous section, the implementation varies according to
the nature of the endogenous variable $X_{2}$ and the instrument $Z_{2}$
(whether continuous or discrete). In the continuous case we need to choose
$\{J_{n},K_{n},\lambda_{n}\}$ for estimating $h_{0}.$ We can make these
choices simultaneously by Generalized Cross-validation (cf. \cite{Wahba_Book},
GCV henceforth). To simplify the computations we implement GCV by setting
first $J_{n}=q_{1}+j_{n}$ for fixed value $j_{n}$ in a small grid (e.g.
$j_{n}\in\{4,5,6,7\}),$ then setting $K_{n}=p_{1}+\left\lfloor cj_{n}%
\right\rfloor ,$ for a grid of values for $c$ in $[1,3],$ where $\left\lfloor
x\right\rfloor $ is the floor function, and then minimizing in $\tau
=\{j_{n},c,\lambda\}$ the GCV criteria $GCV_{n}(\tau)$ given below in
(\ref{GCV}) over the grid values.

Details are given as follows. Let $H_{n}$ denote the $n\times p$ matrix with
rows $\hat{h}_{n}(Z_{i})$ $i=1,...,n.$ Let $\mathbf{X}=[X_{1},...,X_{n}%
]^{\prime}$ and denote by $\mathbf{X}_{1}\equiv\mathbf{Z}_{1}$ and
$\mathbf{X}_{2},$ respectively, the corresponding $n\times p_{1}$ and $n\times
p_{2}$ design matrices based on the partition $X=(X_{1}^{\prime},X_{2}%
^{\prime})^{\prime}$. Construct the $n\times J_{n}$ matrix $Q=[\mathbf{Z}_{1}$
$Q_{2}],$ $J_{n}=q_{1}+j_{n},$ $Q_{2}=[q^{j_{n}}(Z_{21}),...,q^{j_{n}}%
(Z_{2n})]^{\prime}$ ($Q_{2}$ excludes an intercept), and similarly the
$n\times K_{n}$ matrix $P=[\mathbf{X}_{1}$ $P_{2}],$ $K_{n}=p_{1}+k_{n},$
$k_{n}=\left\lfloor cj_{n}\right\rfloor ,$ $P_{2}=[p^{k_{n}}(X_{21}%
),...,p^{k_{n}}(X_{2n})]^{\prime}$ ($P_{2}$ excludes an intercept), and their
corresponding projection matrices $\Pi_{P}$ and $\Pi_{Q},$ where $\Pi
_{A}=A(A^{\prime}A)^{-1}A^{\prime}$ for a generic matrix $A.$ Denote also
$I_{d}$ as the $d\times d$ identity matrix$.$ Then, to provide an expression
for $H_{n}$ we construct%
\[
H_{2n}=Q\hat{A}_{\lambda_{n}}^{-1}Q^{\prime}\Pi_{P}\mathbf{X}_{2},
\]
where%
\[
\hat{A}_{\lambda_{n}}=Q^{\prime}(\Pi_{P}+\lambda_{n}I_{n})Q.
\]
Finally,
\[
H_{n}=[\mathbf{Z}_{1}\text{ }H_{2n}]
\]
and
\begin{equation}
\hat{\beta}=\left(  H_{n}^{\prime}\mathbf{X}\right)  ^{-1}H_{n}^{\prime
}\mathbf{Y}, \label{betahat}%
\end{equation}
where $\mathbf{Y}=[Y_{1},...,Y_{n}]^{\prime}.$ This provides a matrix formula
implementation for our estimator.

To give the GCV criteria define $L_{\tau}=\mathbf{X}\left(  H_{n}^{\prime
}\mathbf{X}\right)  ^{-1}H_{n}^{\prime},$ $\hat{Y}_{\tau}=L_{\tau}%
\mathbf{Y}=(\hat{Y}_{\tau1},...,\hat{Y}_{\tau n})^{\prime}$ and $v_{\tau
}=trace(L_{\tau}).$ Then, the GCV criteria for estimating $\hat{\beta}$ is
\begin{equation}
GCV_{n}(\tau)=\frac{1}{n}\sum_{i=1}^{n}\left(  \frac{Y_{i}-\hat{Y}_{\tau i}%
}{1-(v_{\tau}/n)}\right)  ^{2}. \label{GCV}%
\end{equation}
To estimate $g_{0}$ in the presence of a high dimensional vector of controls
we follow the specification in (\ref{Partlylinear}). The $n\times1$ vector
$G_{n}$ of fitted values $\hat{g}_{n}(X_{i}),$ $i=1,...,n,$ is given by%
\begin{equation}
G_{n}=P\hat{B}_{\lambda_{n}}^{-1}P^{\prime}\Pi_{Q}\mathbf{Y}, \label{Gnnonp}%
\end{equation}
where%
\[
\hat{B}_{\lambda_{n}}=P^{\prime}(\Pi_{Q}+\lambda_{n}I_{n})P.
\]
Since $G_{n}$ is linear in $\mathbf{Y},$ we can easily set another GCV method
for selecting $\{J_{n},K_{n},\lambda_{n}\}$ for $\hat{g}_{n}$ (simply replaced
$L_{\tau}$ above by $L_{\tau}=P\hat{B}_{\lambda_{n}}^{-1}P^{\prime}\Pi_{Q}).$
See also \cite{CFF_2017}$.$

The following algorithm summarizes the main steps for
implementation\footnote{Matlab and R code to implement the TSIV and related
inferences are available from the authors upon request.}:

\begin{description}
\item[Step 1.] Compute $\tau_{n}=\arg\min GCV_{n}(\tau),$ over a finite grid
of values of $\tau=\{j,c,\lambda\}.$

\item[Step 2.] Compute $\hat{\beta}\ $following (\ref{betahat}).

\item[Step 3.] Compute $\hat{g}_{n}$ following (\ref{Gnnonp}).

\item[Step 4.] Compute $\hat{m}_{ni}=m(W_{i},\hat{\beta},\hat{h}_{n},\hat
{g}_{n})$ and $\hat{\Sigma}=E_{n}[\hat{h}_{n}X_{i}^{\prime}]^{-1}E_{n}[\hat
{m}_{ni}\hat{m}_{ni}^{\prime}]E_{n}[X_{i}\hat{h}_{n}^{\prime}]^{-1}.$
\end{description}

For continuous variables we recommend using B-splines as sieve basis. If
$Z_{2}$ is discrete, with support $\{z_{21},...,z_{2j_{2}}\}$, we set
$J_{n}=q_{1}+j_{2}-1$ and $q_{j}(z_{2})=1(z_{2}=z_{2j}),$ $j=2,...,j_{2},$ in
the algorithm above. Similarly, if $X_{2}$ is discrete, with support
$\{x_{21},...,x_{2k_{2}}\}$, we set $K_{n}=p_{1}+k_{2}-1,$ and $p_{k}%
(x_{2})=1(x_{2}=x_{2k}),$ $k=2,...,k_{2}.$ In this discussion, we exclude the
first element in the indicators because the intercept is part of the exogenous controls.

\subsection{Weighted least squares}

\label{WLS}

Our previous discussion can be extended to weighted least squares criteria.
That is, suppose that the OLIVA is now defined as
\begin{subequations}
\begin{equation}
\beta_{w}=\arg\min_{\gamma\in\mathbb{R}^{p}}E[\left(  g(X)-\gamma^{\prime
}X\right)  ^{2}w(X)], \label{BLAw}%
\end{equation}
where $w(X)$ is a positive weight function. This extension can be relevant in
a number of applications. For example, if $f$ is the density of $X$ and
$f^{\ast}$ is a counterfactual density, by taking $w(x)=f^{\ast}(x)/f(x)$ the
linear approximation is under a counterfactual density which might better
summarized the interest of the researcher. Our theory can be extended to this
setting as follows. The necessary condition for regular identification of
$\beta_{w}$ is now%
\end{subequations}
\begin{equation}
E[\left.  h(Z)\right\vert X]=Xw(X)\text{ a.s,} \label{Reg2}%
\end{equation}
for an square integrable $h(\cdot);$ and under this condition and if
$E[XX^{\prime}w(X)]$ is positive definite, then it follows that%
\begin{align*}
\beta_{w}  &  =E[XX^{\prime}w(X)]^{-1}E[Xw(X)g(X)]\\
&  =E[h(Z)X^{\prime}]^{-1}E[h(Z)Y].
\end{align*}
The estimation proceeds as in our basic case (where $w=1)$. If $w$ is unknown,
we can estimate $w$ nonparametrically and use the plugging estimator with the
estimated $w$ to solve for $h$ in (\ref{Reg2}). Our estimator will be
consistent and asymptotically normal under regularity conditions, as in the
basic case. It remains to study if estimation of $w$ changes the asymptotic
variance of the resulting estimator of $\beta_{w}$. This issue is, however,
beyond the scope of this paper and is left for future research.

\section{A Robust Hausman Test}

\label{HausmanTest}

Applied researchers are concerned about the presence of endogeneity, and they
traditionally use tools such as the \cite{Hausman}'s exogeneity test for its
measurement. This test, however, is uninformative under misspecification; see
\cite{Lochner_Moretti_2015}. The reason for this lack of robustness is that in
these cases OLS and IV estimate different objects under exogeneity, with the
estimand of standard IV depending on the instrument itself. As an important
by-product of our analysis, we robustify the classic Hausman test of
exogeneity against nonparametric misspecification of the linear regression model.

The classical Hausman test of exogeneity (cf. \cite{Hausman}) compares OLS
with IV. If we use the TSIV as the IV estimator, we obtain a robust version of
the classical Hausman test, robust to the misspecification of the linear
model. For implementation purposes it is convenient to use a regression-based
test (see \cite{Wooldridge}, pg. 481). We illustrate the idea in the case of
one potentially endogenous variable $X_{2}$ and several exogenous variables
$X_{1}$, with $X_{1}$ including an intercept.

In the model%
\[
Y=\beta_{1}^{\prime}X_{1}+\beta_{2}X_{2}+U,\qquad E[Uh(Z)]=0,\text{
}h(Z)=(X_{1}^{\prime},h_{2}(Z))^{\prime},
\]
the variable $X_{2}$ is exogenous if $Cov(X_{2},U)=0.$ If we write the
first-stage as%
\[
X_{2}=\alpha_{1}^{\prime}X_{1}+\alpha_{2}h_{2}(Z)+V,\qquad E[Vh(Z)]=0,
\]
then weak exogeneity of $X_{2}$ is equivalent to $Cov(V,U)=0.$ This in turn is
equivalent to $\rho=0$ in the least squares regression%
\[
U=\rho V+\xi.
\]
A simple way to run a test for $\rho=0$ is to consider the augmented
regression%
\[
Y=\beta^{\prime}X+\rho V+\xi,
\]
estimated by OLS and use a standard $t-test$ for $\rho=0.$

Since $V$ is unobservable, we first need to obtain residuals from a regression
of the endogenous variable $X_{2}$ on $X_{1}$ and $\hat{h}_{2n}(Z),$ say
$\hat{V}.$ Then, run the regression of $Y$ on $X$ and $\hat{V}$. The new
Hausman test is a standard two-sided t-test for the coefficient of $\hat{V},$
or its Wald version in the multivariate endogenous case. Denote the t-test
statistic by $t_{n}.$ The benefit of this regression approach is that under
some regularity conditions given in Appendix A no correction is necessary in
the OLS standard errors because $\hat{V}$ is estimated. Denote
$S=(X,V)^{\prime},$ and consider the following mild assumption.\bigskip

\noindent\textbf{Assumption 4}: The matrix $E[SS^{\prime}]$ is finite and
non-singular.\bigskip

\begin{theorem}
\label{Hausman} Let Assumptions 1-4 above and Assumptions A1-A6 in the
Appendix A hold. Then, under the the null of exogeneity of $X_{2},$
{$t_{n}\longrightarrow_{d}N(0,1).$}
\end{theorem}

{The proof of Theorem \ref{Hausman} is involved and requires stronger
conditions than that of }Theorem \ref{ANTSIV}. In particular, for obtaining
the result that standard OLS theory applies under the null hypothesis we have
used a conditional exogeneity assumption between $U$ and $Z,$ $E[\left.
U\right\vert Z]=0$ a.s. Simulations below show that, at least for the models
considered, this assumption leads to a robust Hausman test that is able to
control the empirical size. We note that under the null of exogeneity we do
not require the model to be linear in the sense of $E[\left.  U\right\vert
X]=0$ a.s.

\section{Monte Carlo}

\label{MC}

This section studies the finite sample performance of the proposed methods.
Consider the following Data Generating Process (DGP):%
\[
\left\{
\begin{array}
[c]{c}%
Y=\sum_{j=1}^{p}H_{j}(X)+\varepsilon,\\
Z=s(D),\\
\varepsilon=\rho_{\varepsilon}V+\zeta,
\end{array}
\right.  \qquad\left(
\begin{array}
[c]{c}%
X\\
D
\end{array}
\right)  \sim N\left(  \left(
\begin{array}
[c]{c}%
0\\
0
\end{array}
\right)  ,\left(
\begin{array}
[c]{cc}%
1 & \gamma\\
\gamma & 1
\end{array}
\right)  \right)  ,
\]
where $H_{j}(x)$ is the $j-th$ Hermite polynomial, with the first four given
by $H_{0}(x)=1,$ $H_{1}(x)=x,$ $H_{2}(x)=x^{2}-1$ and $H_{3}(x)=x^{3}-3x;$
$V=X-E[\left.  X\right\vert Z],$ $\zeta$ is a standard normal, drawn
independently of $X$ and $D,\ $and $s$ is a monotone function given below. The
DGP is indexed by $p$ and the function $s.$ To generate $V$ note%
\[
E[\left.  X\right\vert Z]=E[\left.  E[\left.  X\right\vert D]\right\vert
Z]=\gamma E[\left.  D\right\vert Z]=\gamma s^{-1}(Z),
\]
where $s^{-1}$ is the inverse of $s.$ Thus, by construction $Z$ is exogenous,
$E[\left.  \varepsilon\right\vert Z]=0,$ while $X$ is endogenous because
$E[\left.  \varepsilon\right\vert X]=\rho X,$ with $\rho=\rho_{\varepsilon
}(1-\gamma^{2})$, $\rho_{\varepsilon}>0$ and $-1<\gamma<1.$

The structural function $g$ is given by
\[
g(x)=\sum_{j=1}^{p}H_{j}(X),
\]
and is therefore linear for $p=1$, but nonlinear for $p>1.$ It follows from
the orthogonality of Hermite polynomials that the true value for OLIVA is
$\beta=1\ $and that $g$ is identified if $\gamma\neq0$ (since $Var(E[\left.
g(X)\right\vert Z])=\sum_{j=1}^{\infty}g_{j}^{2}\gamma^{2j},$ where
$g_{j}=E[g(X)H_{j}(X)]$ is the $j-th$ Hermite coefficient, and thus,
$E[\left.  g(X)\right\vert Z]=0\Longrightarrow g=0$).

Note also that the OLIVA is regularly identified, because $h(Z)=s^{-1}%
(Z)/\gamma$ solves%
\[
E[\left.  h(Z)\right\vert X]=X.
\]
We consider three different DGPs, corresponding to different values of $p$ and
functional forms for $s$:

\begin{description}
\item[DGP1:] $p=1$ and $s(D)=D$ (linear; $s^{-1}(Z)=Z);$

\item[DGP2:] $p=2$ and $s(D)=D^{3}$ (nonlinear; $s^{-1}(Z)=Z^{1/3});$

\item[DGP3:] $p=3$ and $s(D)=\exp(D)/(1+\exp(D))$ (nonlinear; $s^{-1}%
(Z)=\log(Z)-\log(1-Z));$
\end{description}

Several values for the parameters $(\gamma,\rho)$ will be considered:
$\gamma\in\{0.4,0.8\}$ and $\rho\in\{0,0.3,0.9\}$. We will compare the TSIV
with OLS and standard IV (using instrument $Z).$ For DGP1, $h(Z)=\gamma^{-1}Z$
and hence the standard IV estimator with instrument $Z$ is a consistent
estimator for the OLIVA. Indeed, the standard IV can be seen as an oracle
(infeasible version of our TSIV) under DGP1, where $h$ is known rather than
estimated. This allows us to see the effect of estimating $h_{0}$ on
inferences. For DGP2 and DGP3, IV is not consistent for the OLIVA. The number
of Monte Carlo replications is $5000$. The sample sizes considered are
$n=100,$ $500$ and $1000$.

Tables 1-3 report the Bias and MSE for OLS, IV and the TSIV for DGP1-DGP3,
respectively. Our estimator is implemented with B-splines, following the GCV
described in Section \ref{Implementation}, where to simplify the computations
we set $J_{n}=6$ and $K_{n}=2J_{n},$ and optimize only in $\lambda$ for each
simulated data. A similar strategy was followed in \cite{BCK_2007}. Likewise,
we have followed a simple rule for selecting $\{J_{n},K_{n},\lambda_{n}\}$ for
$\hat{g}_{n}$: switch the values of $J_{n}$ and $K_{n}$ used for $\hat{h}_{n}$
to compute $\hat{g}_{n}$ (so now $J_{n}=2K_{n}),$ and use same value of
$\lambda_{n}$ for $\hat{g}_{n}$ as for estimating $\hat{h}_{n},$ which seems
to work well in our simulations. Remarkably, for DGP1 in Table 1 our TSIV
implemented with GCV performs comparably or even better than IV (which does
not estimate $h$ and uses the true $h)$. Thus, our estimator seems to have an
oracle property, performing as well as the method that uses the correct
specification of the model. As expected, OLS is best under exogeneity, but it
leads to large biases under endogeneity. For the nonlinear models DGP2 and
DGP3, IV deteriorates because the linear model is misspecified. Our TSIV
performs well, with a MSE that converges to zero as $n$ increases. Increasing
$\gamma$ makes the instrument stronger, thereby reducing the MSE of IV
estimates, while for a fixed $\gamma,$ increasing the level of endogeneity
increases the MSE.

\begin{table}[ptb]
\caption{Bias and MSE for DGP $1$.}
\centering
\subfloat{
		\scalebox{.8}{
			\begin{tabular}{ccrrrrrrr}
				\hline
				$\rho$ & $\gamma$ & n & BIAS\_OLS & BIAS\_IV & BIAS\_TSIV & MSE\_OLS & MSE\_IV & MSE\_TSIV \\
				\hline
0.0 & 0.4 &  100 & -0.0021 & -0.0019 &  0.0010 & 0.0109 & 0.0829 & 0.0554 \\
&   &  500 &  0.0017 &  0.0025 &  0.0020 & 0.0021 & 0.0127 & 0.0105 \\
&  & 1000 & -0.0001 &  0.0018 &  0.0020 & 0.0010 & 0.0067 & 0.0054 \\
& 0.8 &  100 & -0.0030 & -0.0040 & -0.0040 & 0.0102 & 0.0163 & 0.0159 \\
&  &  500 &  0.0001 & -0.0004 & -0.0004 & 0.0019 & 0.0030 & 0.0030 \\
&  & 1000 &  0.0019 &  0.0025 &  0.0026 & 0.0010 & 0.0016 & 0.0016 \\
0.3 & 0.4 &  100 &  0.2950 & -0.0101 &  0.0841 & 0.0968 & 0.0908 & 0.0729 \\
&  &  500 &  0.2993 &  0.0026 &  0.0347 & 0.0915 & 0.0145 & 0.0168 \\
&  & 1000 &  0.3006 & -0.0003 &  0.0189 & 0.0914 & 0.0071 & 0.0080 \\
& 0.8 &  100 &  0.2956 & -0.0107 &  0.0061 & 0.0987 & 0.0207 & 0.0216 \\
&  &  500 &  0.2991 &  0.0009 &  0.0038 & 0.0918 & 0.0039 & 0.0039 \\
&  & 1000 &  0.2987 & -0.0023 & -0.0012 & 0.0904 & 0.0019 & 0.0019 \\
0.9 & 0.4 &  100 &  0.8993 & -0.0827 &  0.1753 & 0.8213 & 0.1990 & 0.1569 \\
&  &  500 &  0.9028 & -0.0145 &  0.0421 & 0.8173 & 0.0295 & 0.0296 \\
&  & 1000 &  0.8998 & -0.0066 &  0.0231 & 0.8108 & 0.0130 & 0.0140 \\
& 0.8 &  100 &  0.8965 & -0.0186 &  0.0287 & 0.8270 & 0.0573 & 0.0571 \\
&  &  500 &  0.8980 & -0.0036 &  0.0030 & 0.8114 & 0.0108 & 0.0109 \\
&  & 1000 &  0.8993 &  0.0031 &  0.0058 & 0.8111 & 0.0049 & 0.0050 \\
				\hline
			\end{tabular}
		}}
\end{table}

\begin{table}[ptb]
\caption{Bias and MSE for DGP $2$.}
\centering
\subfloat{
			\scalebox{.8}{
				\begin{tabular}{ccrrrrrrr}
					\hline
					$\rho$ & $\gamma$ & n & BIAS\_OLS & BIAS\_IV & BIAS\_TSIV & MSE\_OLS & MSE\_IV & MSE\_TSIV \\
					\hline
0.0 & 0.4 &  100 &  0.0131 & -0.0030 & -0.0037 & 0.1009 & 0.6321 & 0.2226 \\
&  &  500 &  0.0083 &  0.0216 &  0.0126 & 0.0213 & 0.1319 & 0.0479 \\
&  & 1000 &  0.0021 &  0.0005 &  0.0034 & 0.0115 & 0.0764 & 0.0228 \\
& 0.8 &  100 & -0.0012 &  0.0001 & -0.0001 & 0.0990 & 0.4559 & 0.1286 \\
&  &  500 &  0.0015 &  0.0056 &  0.0032 & 0.0211 & 0.1261 & 0.0275 \\
&  & 1000 &  0.0019 &  0.0084 &  0.0030 & 0.0113 & 0.0689 & 0.0154 \\
0.3 & 0.4 &  100 &  0.2932 & -0.0472 &  0.0605 & 0.1859 & 0.6167 & 0.2342 \\
&  &  500 &  0.2874 & -0.0325 &  0.0302 & 0.1023 & 0.1417 & 0.0594 \\
&  & 1000 &  0.3008 & -0.0135 &  0.0402 & 0.1013 & 0.0778 & 0.0331 \\
& 0.8 &  100 &  0.3064 &  0.0083 &  0.0318 & 0.1987 & 0.4554 & 0.1400 \\
&  &  500 &  0.3020 &  0.0078 &  0.0208 & 0.1114 & 0.1226 & 0.0289 \\
&  & 1000 &  0.3046 &  0.0076 &  0.0248 & 0.1040 & 0.0647 & 0.0168 \\
0.9 & 0.4 &  100 &  0.9053 & -0.1359 &  0.2155 & 0.9270 & 1.0165 & 0.3615 \\
&  &  500 &  0.8968 & -0.0093 &  0.0794 & 0.8260 & 0.1619 & 0.0914 \\
&  & 1000 &  0.8974 & -0.0122 &  0.0493 & 0.8159 & 0.0817 & 0.0449 \\
& 0.8 &  100 &  0.9095 & -0.0117 &  0.0491 & 0.9425 & 0.5482 & 0.1921 \\
&  &  500 &  0.8969 & -0.0013 &  0.0226 & 0.8290 & 0.1405 & 0.0435 \\
&  & 1000 &  0.8981 & -0.0021 &  0.0271 & 0.8185 & 0.0753 & 0.0220 \\
					\hline
				\end{tabular}
			}}
\end{table}

\begin{table}[ptb]
\caption{Bias and MSE for DGP $3$.}
\centering
\subfloat{
				\scalebox{.8}{
					\begin{tabular}{ccrrrrrrr}
						\hline
						$\rho$ & $\gamma$ & n & BIAS\_OLS & BIAS\_IV & BIAS\_TSIV & MSE\_OLS & MSE\_IV & MSE\_TSIV \\
						\hline
0.0 & 0.4 &  100 & -0.0570 & -1.5268 & -0.0717 & 0.5023 &   381.7332 & 0.6817 \\
&  &  500 & -0.0021 & -0.5039 & -0.0346 & 0.1000 &   155.9296 & 0.1326 \\
&  & 1000 & -0.0014 & -0.0365 & -0.0378 & 0.0550 &     0.6179 & 0.0681 \\
& 0.8 &  100 & -0.0418 & -0.4112 & -0.1106 & 0.4795 &     2.6703 & 0.4935 \\
&  &  500 & -0.0096 & -0.2270 & -0.0411 & 0.1072 &     0.4192 & 0.1084 \\
&  & 1000 & -0.0113 & -0.2150 & -0.0330 & 0.0527 &     0.2452 & 0.0543 \\
0.3 & 0.4 &  100 &  0.2899 & -5.4825 &  0.0227 & 0.6475 & 28179.2626 & 0.8182 \\
&  &  500 &  0.2882 & -0.1335 &  0.0060 & 0.1878 &     1.5707 & 0.1571 \\
&  & 1000 &  0.2887 & -0.0822 &  0.0199 & 0.1351 &     0.6518 & 0.0926 \\
& 0.8 &  100 &  0.2693 & -0.3815 & -0.0857 & 0.5906 &    11.1463 & 0.5498 \\
&  &  500 &  0.3062 & -0.1985 & -0.0249 & 0.2061 &     0.4885 & 0.1221 \\
&  & 1000 &  0.2951 & -0.2166 & -0.0246 & 0.1395 &     0.2512 & 0.0570 \\
0.9 & 0.4 &  100 &  0.8470 &  1.4445 &  0.1675 & 1.1993 &  1772.3946 & 0.8970 \\
&  &  500 &  0.8888 & -0.3336 &  0.0449 & 0.9098 &     4.8599 & 0.2103 \\
&  & 1000 &  0.8914 & -0.1313 &  0.0158 & 0.8473 &     0.8558 & 0.0982 \\
& 0.8 &  100 &  0.8341 & -0.5724 & -0.0917 & 1.1833 &     4.3735 & 0.6045 \\
&  &  500 &  0.8749 & -0.2933 & -0.0566 & 0.8668 &     0.6084 & 0.1301 \\
&  & 1000 &  0.8863 & -0.2466 & -0.0401 & 0.8380 &     0.2861 & 0.0681 \\
						\hline
					\end{tabular}
				}}
\end{table}

We have done extensive sensitivity analysis on the performance of the TSIV
estimator. Simulations in the Supplemental Appendix report the sensitivity of
the estimator to different choices of tuning parameters, $J_{n}$, $K_{n}$ and
$\lambda_{n}$. From these results, we see that the TSIV estimator is not
sensitive to the choice of these parameters, within the wide ranges for which
we have experimented. This is consistent with the regular identification,
which means that the estimator should be robust to local perturbations of the
tuning parameters. Likewise, unreported simulations with other DGPs confirm
the overall good performance of the proposed TSIV under different scenarios.

Table 4 provides the results for coverage of confidence intervals based on the
asymptotic normality of the TSIV using the GCV-computed $\lambda_{n}$, along
with that using $0.7\lambda_{n}$ and $0.9\lambda_{n}$. The coverage is very
stable for the three choices of $\lambda_{n}$ considered. The performance in
DGP1 and DGP2 is fairly good, while in DGP3 it noticeably improves when the
sample size increases.

\begin{table}[ptb]
\caption{$95\%$ coverage for TSIV.}%
\centering
\scalebox{.8}{
\begin{tabular}{cccrrrrrrrrr}
	\hline
\multicolumn{3}{c}{} & \multicolumn{3}{c}{DGP1} & \multicolumn{3}{c}{DGP2} &\multicolumn{3}{c}{DGP3}    \\
\hline
	$\rho$ & $\gamma$ & n & 0.7cv & 0.9cv & 1.0cv & 0.7cv & 0.9cv& 1.0cv & 0.7cv & 0.9cv & 1.0cv \\
	\hline
	0.0 & 0.4 &  100 & 0.973 & 0.976 & 0.976 & 0.950 & 0.954 & 0.955 & 0.899 & 0.901 & 0.903 \\
	 &  &  500 & 0.976 & 0.978 & 0.977 & 0.950 & 0.951 & 0.951 & 0.929 & 0.931 & 0.932 \\
	 &  & 1000 & 0.971 & 0.973 & 0.973 & 0.954 & 0.957 & 0.956 & 0.931 & 0.931 & 0.930 \\
	 & 0.8 &  100 & 0.964 & 0.965 & 0.966 & 0.929 & 0.929 & 0.931 & 0.837 & 0.837 & 0.838 \\
	 &  &  500 & 0.957 & 0.957 & 0.957 & 0.941 & 0.942 & 0.944 & 0.902 & 0.905 & 0.905 \\
	 &  & 1000 & 0.950 & 0.951 & 0.951 & 0.932 & 0.938 & 0.941 & 0.926 & 0.927 & 0.927 \\
	0.3 & 0.4 &  100 & 0.976 & 0.982 & 0.982 & 0.950 & 0.948 & 0.949 & 0.919 & 0.921 & 0.922 \\
	 &  &  500 & 0.957 & 0.957 & 0.959 & 0.949 & 0.952 & 0.950 & 0.931 & 0.933 & 0.932 \\
	 &  & 1000 & 0.964 & 0.965 & 0.965 & 0.938 & 0.939 & 0.938 & 0.936 & 0.936 & 0.934 \\
	 & 0.8 &  100 & 0.945 & 0.945 & 0.946 & 0.917 & 0.920 & 0.920 & 0.858 & 0.861 & 0.862 \\
	 &  &  500 & 0.944 & 0.941 & 0.941 & 0.946 & 0.946 & 0.946 & 0.917 & 0.920 & 0.921 \\
	 &  & 1000 & 0.961 & 0.960 & 0.960 & 0.940 & 0.941 & 0.941 & 0.917 & 0.923 & 0.923 \\
	0.9 & 0.4 &  100 & 0.903 & 0.901 & 0.902 & 0.938 & 0.943 & 0.943 & 0.955 & 0.957 & 0.956 \\
	 &  &  500 & 0.947 & 0.949 & 0.948 & 0.936 & 0.940 & 0.941 & 0.951 & 0.949 & 0.949 \\
	 &  & 1000 & 0.943 & 0.942 & 0.942 & 0.925 & 0.929 & 0.932 & 0.950 & 0.951 & 0.951 \\
	 & 0.8 &  100 & 0.931 & 0.930 & 0.930 & 0.920 & 0.921 & 0.921 & 0.899 & 0.898 & 0.898 \\
	 &  &  500 & 0.938 & 0.937 & 0.935 & 0.949 & 0.949 & 0.949 & 0.918 & 0.920 & 0.921 \\
	    &  & 1000 & 0.951 & 0.951 & 0.951 & 0.954 & 0.954 & 0.954 & 0.930 & 0.935 & 0.935 \\
	\hline
\end{tabular}
}\end{table}

We now turn to the Hausman test. Practitioners often use the Hausman test to
empirically evaluate the presence of endogeneity. As mentioned above, the
standard Hausman test is not robust to misspefication of the linear model,
because in that case OLS and IV estimate different parameters
(\cite{Lochner_Moretti_2015}). We confirm this by simulating data from
DGP1-DGP3 and reporting rejection frequencies for the standard Hausman test
for $\gamma\in\{0.4,0.8\}$. Table 5 contains the results. For DGP1, the
rejection frequencies for $\rho=0$ are close to the nominal level of 5\%
across the different sample sizes, confirming the validity of the test under
correct specification. However, for DGP2 and DGP3 we observe large size
distortions for the standard Hausman test, as large as 85\%. This shows that
the standard Hausman test is unreliable under misspecification of the linear
model. In contrast, the proposed robust tests is able to control type-I error
uniformly across the three DGPs. We also report size-corrected empirical
rejections under the alternative. For the linear model, the standard Hausman
test has (slightly) larger power than the robust test, while for the nonlinear
model DGP2, the robust test has much larger power. For DGP3, the robust
Hausman test outperforms the standard test for low values of $\rho,$ while for
large values of $\rho$ they have comparable powers. In all cases we observe an
empirical power that increases with the sample size and the level endogeneity,
suggesting consistency against these alternatives. Despite these simulation
results and others in the Supplemental Appendix, we stress that standard and
robust Hausman tests should be viewed as complements rather than substitutes,
given that they work under different set of assumptions.

\begin{table}[ptb]
\caption{Empirical size and size-corrected power for Standard (S) and Robust
(R) Hausman tests.}%
\centering
\scalebox{.8}{
\begin{tabular}{ccccccccc}
	\hline
	\multicolumn{3}{c}{} & \multicolumn{2}{c}{DGP1} & \multicolumn{2}{c}{DGP2} &\multicolumn{2}{c}{DGP3}    \\
\hline
$\rho$ & $\gamma$ & n & S & R & S & R & S & R \\
\hline
0.0 & 0.4 &  100 & 0.068 & 0.051 & 0.142 & 0.038 & 0.054 & 0.016 \\
&  &  500 & 0.062 & 0.044 & 0.070 & 0.016 & 0.048 & 0.010 \\
&  & 1000 & 0.056 & 0.040 & 0.053 & 0.006 & 0.050 & 0.008 \\
& 0.8 &  100 & 0.071 & 0.063 & 0.221 & 0.015 & 0.090 & 0.001 \\
&  &  500 & 0.047 & 0.045 & 0.146 & 0.004 & 0.521 & 0.004 \\
&  & 1000 & 0.060 & 0.054 & 0.110 & 0.001 & 0.850 & 0.002 \\
0.3 & 0.4 &  100 & 0.242 & 0.174 & 0.066 & 0.082 & 0.085 & 0.105 \\
&  &  500 & 0.802 & 0.718 & 0.174 & 0.292 & 0.243 & 0.284 \\
&  & 1000 & 0.985 & 0.950 & 0.296 & 0.549 & 0.414 & 0.492 \\
& 0.8 &  100 & 0.952 & 0.896 & 0.108 & 0.562 & 0.671 & 0.727 \\
&  &  500 & 1.000 & 0.999 & 0.216 & 0.924 & 0.982 & 1.000 \\
&  & 1000 & 1.000 & 1.000 & 0.340 & 0.942 & 0.992 & 1.000 \\
0.9 & 0.4 &  100 & 0.958 & 0.754 & 0.281 & 0.386 & 0.412 & 0.418 \\
&  &  500 & 1.000 & 0.993 & 0.744 & 0.952 & 0.930 & 0.938 \\
&  & 1000 & 1.000 & 1.000 & 0.932 & 0.994 & 0.999 & 0.997 \\
& 0.8 &  100 & 1.000 & 0.992 & 0.370 & 0.956 & 1.000 & 0.998 \\
&  &  500 & 1.000 & 1.000 & 0.739 & 0.980 & 1.000 & 1.000 \\
&  & 1000 & 1.000 & 1.000 & 0.910 & 0.980 & 1.000 & 1.000 \\
\hline
\end{tabular}
}\end{table}

We also report in the Supplemental Appendix further simulation results for
cases where $Z$ is discrete and $X$ is continuous. For these DGPs $g$ is not
identified, although Assumption 3 is satisfied. These additional simulation
results provide further evidence of the excellent finite sample performance of
the TSIV and the robust Hausman test relative to their standard IV counterparts.

Overall, these simulations confirm the robustness of the proposed methods to
misspecification of the linear IV model and their adaptive behaviour when
correct specification holds. Furthermore, the TSIV estimator does not seem to
be too sensitive to the choice of tuning parameters. Finally, the proposed
Hausman test is indeed robust to the misspecification of the linear model,
which makes it a reliable tool for economic applications. These finite sample
robustness results confirm the claims made for the TSIV estimator as a
nonparametric analog to OLS under endogeneity.

\section{Estimating the Elasticity of Intertemporal Substitution}

\label{EIS}

In its log-linearized version, the Consumption-based Capital Asset Pricing
Model (CCAPM) leads to the equation%
\begin{equation}
\Delta c_{t+1}=\alpha+\psi r_{t+1}+U_{t},\qquad E[U_{t}\mid Z_{t}]=0\text{
a.s.,} \label{reg1}%
\end{equation}
where $\psi$ is the elasticity of intertemporal substitution (EIS), $\Delta
c_{t+1}$ is the growth rate of consumption (the first difference in log real
consumption per capita), $r_{t+1}$ is the real interest rate at time $t+1$,
$\alpha$ is a constant and $Z_{t}$ is a vector of variables in the agent's
information set at time $t$. The parameters $\beta_{0}=(\alpha,\psi)^{\prime}$
can be estimated from (\ref{reg1}) by several estimation strategies; see,
e.g., \cite{HansenSingleton_1983}. \cite{Yogo}, using data from
\cite{Campbell}, applied Two-Step Least Squares (TSLS), among other methods,
to obtain estimates of $\psi$ across different countries, arguing that in most
cases the TSLS is subject to weak identification. Here we focus on quarterly
US interest rate data, for which there is empirical evidence suggesting
identification (the first-stage F statistic is 15.5). The data set is
available at Motohiro Yogo's web page. A full description of the data is given
in \cite{Campbell}.\footnote{It should be possible to extend our asymptotic
results above to strictly stationary and ergodic time series data, although
doing so is beyond the scope of this paper. Following much of the literature,
including \cite{Yogo}, we compute standard errors assuming that the influence
functions of the reported estimators are uncorrelated.}

Following \cite{Yogo}, we use as instruments $Z_{t}=(r_{t-1},\pi_{t-1},\Delta
c_{t-1},dp_{t-1}),$ where $r_{t}$ is the nominal interest rate, $\pi_{t}$ is
inflation, and $dp_{t}$ is the log dividend-price ratio. The sample size is
$n=206.$ The TSLS point estimate of $\psi$ is 0.06, with a standard error of
0.09. We compare the TSLS with the proposed TSIV. To deal with the curse of
dimensionality, we estimate $h_{0}(Z)$ with an additive nonparametric model,
$h_{0}(Z_{t})=h_{01}(Z_{t1})+\cdots+h_{04}(Z_{t4})$. Specifically, we follow
the implementation in our Monte Carlo and use B-splines with $4,5$ or $6$
knots for each instrument, leading to $J_{n}=12,15,18$, respectively,
$K_{n}=2J_{n}$ and GCV for choosing $\lambda$. The matrix $Q=[q^{J_{n}}%
(Z_{1}),...,q^{J_{n}}(Z_{n})]^{\prime}$ simply concatenates the corresponding
matrix for each instrument. For estimating $g_{0}$ for the TSIV's standard
errors we choose the same $J_{n}$ and $K_{n}$ as before, and compute
$\lambda_{n}$ by GCV. Further details on implementation are given in Section
\ref{Implementation} for the case with an intercept (so $X_{1}=Z_{1}\equiv1).$

\begin{table}[ptb]
\caption{EIS for Quarterly US data}%
\centering\scalebox{.8}{
\begin{tabular}
[c]{cccccc}\hline
& OLS & TSLS &  & TSIV & \\
$J_{n}$ &  &  & 4 & 5 & 6\\\hline
estimate & 0.161 & 0.060 & 0.151 & 0.160 & 0.162\\
s.e. & (0.054) & (0.095) & (0.092) & (0.099) & (0.101)\\
Hausman p-value &  & 0.154 & 0.897 & 0.995 & 0.984\\
	\hline
\end{tabular}
}\end{table}

\noindent Not surprisingly, our TSLS coincides with that of \cite{Yogo}. The
TSIV is relatively much larger than the TSLS, and closer to the OLS, while the
standard errors of both IV methods are similar in magnitude. These results are
robust to the choice of $J_{n}$ and $K_{n}$ (we have experimented with
$K_{n}=cJ_{n}$ for $c$ between 1 and 3 and obtain qualitatively the same
conclusions). If we apply our robust Hausman test of exogeneity we obtain very
large p-values. Again, this result is robust to the choice of $J_{n}$ and
$K_{n}.$ In contrast, the standard Hausman test leads to a p-value of 0.154.

We reach several conclusions from these results. First, the difference between
the TSLS and the TSIV suggests that nonlinearities might be important in this
application (indeed, the plot of the estimated $\hat{g}_{n},$ which is not
reported here for the sake of space, reveals a marked nonlinear estimate).
Second, once one accounts for the misspecification uncertainty, the null
hypothesis of exogeneity cannot be rejected, thereby suggesting that for the
purpose of estimating a log-linearized version of the Euler equation,
endogeneity bias may be a second-order concern.\newpage

\section{Appendix A: Notation, Assumptions and Preliminary Results}

\label{AppendixA}

\subsection{Notation}

Define the kernel subspace $\mathcal{N}\equiv\{f\in L_{2}(X):T^{\ast}f=0\}$ of
the operator $T^{\ast}f(z):=E[\left.  f(X)\right\vert Z=z]$. Let
$Ts(x):=E[\left.  s(Z)\right\vert X=x]$ denote the adjoint operator of
$T^{\ast}$ and let $\mathcal{R}(T):=\{f\in L_{2}(X):\exists s\in
L_{2}(Z),Ts=f\}$ its range. For a subspace $V,$ $V^{\perp},$ $\overline{V}$
and $P_{\overline{V}}$ denote, respectively, its orthogonal complement, its
closure and its orthogonal projection operator. Let $\otimes$ denote Kronecker
product and let $I_{p}$ denote the identity matrix of order $p.$

Define the Sobolev norm $\left\Vert \cdot\right\Vert _{\infty,\eta}$ as
follows. Define for any vector $a$ of $p$ integers the differential operator
$\partial_{x}^{a}:=\partial^{\left\vert a\right\vert _{1}}/\partial
x_{1}^{a_{1}}\ldots\partial x_{p}^{a_{p}},$ where $\left\vert a\right\vert
_{1}:=\sum_{i=1}^{p}a_{i}$. Let $\mathcal{X}$ denote a finite union of convex,
bounded subsets of $\mathbb{R}^{p}$, with non-empty interior. For any smooth
function $h:\mathcal{X}\subset\mathbb{R}^{p}\rightarrow\mathbb{R}$ and some
$\eta>0$, let \underline{$\eta$} be the largest integer smaller than $\eta$,
and%
\[
\left\Vert h\right\Vert _{\infty,\eta}:=\underset{\left\vert a\right\vert
_{1}\leq\underline{\eta}}{\max}\text{ }\underset{x\in\mathcal{X}}{\sup
}\left\vert \partial_{x}^{a}h(x)\right\vert +\underset{\left\vert a\right\vert
_{1}=\underline{\eta}}{\max}\text{ }\underset{x\neq x^{\prime}}{\sup}%
\frac{\left\vert \partial_{x}^{a}h(x)-\partial_{x}^{a}h(x^{\prime})\right\vert
}{\left\vert x-x^{\prime}\right\vert ^{\eta-\underline{\eta}}}\text{.}%
\]
Let $\mathcal{H}$ denote the parameter space for $h,$ and define the
identified set $\mathcal{H}_{0}=\{h\in\mathcal{H}:m(X,h)=0$ a.s.$\}$. The
operator $Th(x):=E[\left.  h(Z)\right\vert X=x]$ is estimated by
\[
\hat{T}h(x):=\hat{E}[\left.  h(Z)\right\vert X=x]=\sum_{i=1}^{n}\left(
{p^{K_{n}}}^{\prime}(x)(P^{\prime}P)^{-1}p^{K_{n}}(X_{i})\otimes
h(Z_{i})\right)  .
\]
The operator $\hat{T}$ is considered as an operator from $\mathcal{H}_{n}$ to
$\mathcal{G}_{n}\subseteq L_{2}(X),$ where $\mathcal{G}_{n}$ is the linear
span of $\{{p^{K_{n}}}(\cdot)\}$. Let $E_{n}[g(W)]$ denote the sample mean
operator, i.e. $E_{n,W}[g(W)]=n^{-1}\sum_{i}^{n}g(W_{i}),$ let $||g||_{n,W}%
^{2}=E_{n}[\left\vert g(W)\right\vert ^{2}]$, and let $\left\langle
f,g\right\rangle _{n,W}=n^{-1}\sum_{i=1}^{n}f(W_{i})g(W_{i})$ be the empirical
$L_{2}$ inner product. We drop the dependence on $W$ for simplicity of
notation. Denote by $\hat{T}^{\ast}$ the adjoint operator of $\hat{T}$ with
respect to the empirical inner product. Simple algebra shows for $p=1,$%
\begin{align*}
\left\langle \hat{T}h,g\right\rangle _{n}  &  =n^{-1}\sum_{i=1}^{n}%
h(Z_{i}){p^{K_{n}}}^{\prime}(X_{i})(P^{\prime}P)^{-1}\sum_{j=1}^{n}p^{K_{n}%
}(X_{j})g(X_{j})\\
&  =\left\langle h,\hat{T}^{\ast}g\right\rangle _{n},
\end{align*}
so $\hat{T}^{\ast}g=P_{\mathcal{H}_{n}}\hat{E}[\left.  g(X)\right\vert
X=\cdot]=P_{\mathcal{H}_{n}}\hat{T}g.$ A similar expression holds for $p>1.$

With this operator notation, the first-step has the expression (where $I$
denotes the identity operator)
\begin{equation}
\hat{h}_{n}=\left(  \hat{T}^{\ast}\hat{T}+\lambda_{n}I\right)  ^{-1}\hat
{T}^{\ast}\hat{X}, \label{Tikh}%
\end{equation}
where $\hat{X}=\hat{E}[\left.  X\right\vert X=\cdot].$ Similarly, define the
Tikhonov approximation of $h_{0}$%
\begin{equation}
h_{\lambda_{n}}=A_{\lambda_{n}}^{-1}T^{\ast}X, \label{App}%
\end{equation}
where $A_{\lambda_{n}}=T^{\ast}T+\lambda_{n}I.$ Define also $B_{\lambda_{n}%
}=TT^{\ast}+\lambda_{n}I.$ With some abuse of notation, denote the operator
norm by%
\[
\left\Vert T\right\Vert =\sup_{h\in\mathcal{H},\left\Vert h\right\Vert \leq
1}\left\Vert Th\right\Vert .
\]
Let $\mathcal{G}\subseteq L_{2}(X)$ denote the parameter space for $g.$ An
envelop for $\mathcal{G}$ is a function $G$ such that $\left\vert
g(x)\right\vert \leq G(x)$ for all $g\in\mathcal{G}.$ Given two functions
$l,u,$ a bracket $[l,u]$ is the set of functions $f\in\mathcal{G}$ such that
$l\leq f\leq u$. An $\varepsilon$-bracket with respect to $\left\Vert
\cdot\right\Vert $ is a bracket $[l,u]$ with $\left\Vert l-u\right\Vert
\leq\varepsilon,$ $\left\Vert l\right\Vert <\infty$ and $\left\Vert
u\right\Vert <\infty$ (note that $u$ and $l$ not need to be in $\mathcal{G}$).
The \textit{covering number with bracketing} $N_{[\cdot]}(\varepsilon
,\mathcal{G},\left\Vert \cdot\right\Vert )$ is the minimal number of
$\varepsilon$-brackets with respect to $\left\Vert \cdot\right\Vert $ needed
to cover $\mathcal{G}$. Define the bracketing entropy%
\[
J_{[\cdot]}(\delta,\mathcal{G},\left\Vert \cdot\right\Vert )=\int_{0}^{\delta
}\sqrt{\log N_{[\cdot]}(\varepsilon,\mathcal{G},\left\Vert \cdot\right\Vert
)}d\varepsilon
\]
Similarly, we define $J_{[\cdot]}(\delta,\mathcal{H},\left\Vert \cdot
\right\Vert ).$ Finally, throughout $C$ denotes a positive constant that may
change from expression to expression.

{Let }$W=(Y,X^{\prime},Z^{\prime})^{\prime}${ be a random vector defined on a
probability space $\left(  \Omega,\mathcal{B},\mathbb{P}\right)  $}. {For a
measurable function $f$ we denote $\mathbb{P}f:=\int fd\mathbb{P},$
\[
\mathbb{P}_{n}f:=\frac{1}{n}%
{\displaystyle\sum\limits_{i=1}^{n}}
f\left(  W_{i}\right)  \text{ and }\mathbb{G}_{n}f:=\sqrt{n}\left(
\mathbb{P}_{n}f-\mathbb{P}f\right)  .
\]
}\bigskip

\subsection{Assumptions}

The following assumptions are standard in the literature of sieve estimation;
see, e.g., \cite{Newey_1997}, \cite{Chen_hb}, \cite{Santos_2011}, and
\cite{Chen_Pouzo_2012}.\bigskip

\noindent\textbf{Assumption A1}: (i) $\{Y_{i},X_{i},Z_{i}\}_{i=1}^{n}$ is an
$iid$ sample, satisfying (\ref{1}) with $E[\left.  \varepsilon\right\vert
Z]=0$ a.s and $E[Y^{2}]<\infty;$ (ii) $X$ has a compact support with
$E[\left\vert X\right\vert ^{2}]<\infty$; (iii) $Z$ has a compact support;
(iv) the densities of $X$ and $Z$ are bounded and bounded away from
zero.\bigskip

\noindent\textbf{Assumption A2}: (i) The eigenvalues of $E[p^{K_{n}%
}(X)p^{K_{n}}(X)^{\prime}]$ are bounded above and away from zero; (ii)
$\max_{1\leq k\leq K_{n}}\left\Vert p_{k}\right\Vert \leq C$ and $\xi
_{n,p}^{2}K_{n}=o(n),$ for $\xi_{n,p}=\sup_{x}\left\vert p^{K_{n}%
}(x)\right\vert ;$ (iii) there is $\pi_{n,p}(h)$ such that $\sup
_{h\in\mathcal{H}}\left\Vert E[\left.  h(Z)\right\vert X=\cdot]-\pi
_{n,p}^{\prime}(h)p^{K_{n}}(\cdot)\right\Vert =O(K_{n}^{-\alpha_{T}});$ (iv)
there is a finite constant $C$, such that $\sup_{h\in\mathcal{H},\left\Vert
h\right\Vert \leq1}\left\vert h(Z)-E[\left.  h(Z)\right\vert X]\right\vert
\leq\rho_{n,p}(Z,X)$ with $E[\left.  \left\vert \rho_{n,p}(Z,X)\right\vert
^{2}\right\vert X]\leq C$.\bigskip

\noindent\textbf{Assumption A3}: (i) The eigenvalues of $E[q^{J_{n}%
}(Z)q^{J_{n}}(Z)^{\prime}]$ are bounded above and away from zero; (ii) there
is a sequence of closed subsets satisfying $\mathcal{H}_{j}\subseteq
\mathcal{H}_{j+1}\subseteq\mathcal{H}$, $\mathcal{H}$ is closed, bounded and
convex, $h_{0}\in\mathcal{H}_{0},$ and there is a $\Pi_{n}(h_{0}%
)\in\mathcal{H}_{n}$ such that $\left\Vert \Pi_{n}(h_{0})-h_{0}\right\Vert
=o(1);$ (iii) $\sup_{h\in\mathcal{H}_{n}}\left\vert \left\Vert h\right\Vert
_{n}^{2}-\left\Vert h\right\Vert ^{2}\right\vert =o_{P}(1);$ (iv) $\lambda
_{n}\downarrow0$ and $\max\{\left\Vert \Pi_{n}(h_{0})-h_{0}\right\Vert
^{2},c_{n,T}^{2}\}=o(\lambda_{n}),$ where $c_{n,T}=\sqrt{K_{n}/n}%
+K_{n}^{-\alpha_{T}};$ (v) $A_{\lambda_{n}}$ is non-singular.\bigskip

\noindent\textbf{Assumption A4}: (i) $h_{0}\in\mathcal{R}(\left(  T^{\ast
}T\right)  ^{\alpha_{h}/2})$ and $g_{0}\in\mathcal{R}(\left(  TT^{\ast
}\right)  ^{\alpha_{g}/2}),$ $\alpha_{h},\alpha_{g}>0;$ (ii) $\max_{1\leq
j\leq J_{n}}\left\Vert q_{j}\right\Vert \leq C$ and $\xi_{n,j}^{2}J_{n}=o(n),$
for $\xi_{n,j}=\sup_{z}\left\vert q^{J_{n}}(z)\right\vert ;$ (iii) $\sup
_{g\in\mathcal{G}}\left\Vert E[\left.  g(X)\right\vert Z=\cdot]-\pi
_{n,q}^{\prime}(g)q^{J_{n}}(\cdot)\right\Vert =O(J_{n}^{-\alpha_{T^{\ast}}%
})\ $for some $\pi_{n,q}(g);$ (iv) $\sup_{g\in\mathcal{G},\left\Vert
g\right\Vert \leq1}\left\vert g(X)-E[\left.  g(X)\right\vert Z]\right\vert
\leq\rho_{n,q}(Z,X)$ with $E[\left.  \left\vert \rho_{n,q}(Z,X)\right\vert
^{2}\right\vert Z]\leq C$; (v) $\lambda_{n}^{-1}c_{n}=o(1),$ where
$c_{n}=c_{n,T}+c_{n,T^{\ast}}$ and $c_{n,T^{\ast}}=\sqrt{J_{n}/n}%
+J_{n}^{-\alpha_{T^{\ast}}};$ (vi) $B_{\lambda_{n}}$ is non-singular.\bigskip

\noindent\textbf{Assumption A5}: (i) $E[\left.  U^{2}\right\vert Z]<C$ a.s.;
(ii) $N_{[\cdot]}(\delta,\mathcal{G},\left\Vert \cdot\right\Vert )<\infty$ and
$J_{[\cdot]}(\delta,\mathcal{H},\left\Vert \cdot\right\Vert )<\infty$ for some
$\delta>0,$ and $\mathcal{G}$ and $\mathcal{H}$ have squared integrable
envelopes$.$\bigskip

\noindent\textbf{Assumption A6}: (i) $\lambda_{n}^{-1}c_{n}=o(n^{-1/4})$; (ii)
$\sqrt{n}\lambda_{n}^{\min(\alpha_{h},2)}=o(1)$ and $\sqrt{n}c_{n}\lambda
_{n}^{\min(\alpha_{h}-1,1)}=o(1);$ (iii) $h_{0}\in\mathcal{R}(T^{\ast}),$
$E\left[  \left.  \left\vert X-h_{0}(Z)\right\vert ^{4}\right\vert X\right]  $
is bounded and $Var[\left.  h_{0}(Z)\right\vert X]$ is bounded and bounded
away from zero; and (iv) $E[\left.  U\right\vert Z]=0$ a.s.\bigskip

\noindent For regression splines $\xi_{n,p}^{2}=O(K_{n}),$ and hence A2(ii)
requires $K_{n}^{2}/n\rightarrow0,$ see \cite{Newey_1997}. Assumptions
A2(iii-iv) are satisfied if $\sup_{h\in\mathcal{H}}\left\Vert Th\right\Vert
_{\infty,\eta_{h}}<\infty$ with $\alpha_{T}=\eta_{h}/q.$ Assumption A3(iii)
holds under mild conditions if for example $\sup_{h\in\mathcal{H}}\left\Vert
h\right\Vert <C.$ Assumption A4(i) is a regularity condition that is well
discussed in the literature, see e.g. \cite{FJV_2011}. A sufficient condition
for Assumption A5(ii) is that for some $\eta_{h}>q/2$ and $\eta_{g}>p/2$ we
have $\sup_{h\in\mathcal{H}}\left\Vert h\right\Vert _{\infty,\eta_{h}}<\infty$
and $\sup_{g\in\mathcal{G}}\left\Vert g\right\Vert _{\infty,\eta_{g}}<\infty;$
see Theorems 2.7.11 and 2.7.1 in \cite{van_der_Vaart_Wellner_Book}.
Assumptions A6 is standard.

\subsection{Preliminary Results}

In all the preliminary results Assumptions 1-3 in the text are assumed to
hold. \bigskip

\noindent\textbf{Lemma A1}: Let Assumptions A1-A3 hold. Then, $\left\Vert
\hat{h}_{n}-h_{0}\right\Vert =o_{P}(1).$

\noindent\textbf{Proof of Lemma A1}: We proceed to verify the conditions of
Theorem A.1 in \cite{Chen_Pouzo_2012}. Recall $\mathcal{H}_{0}=\{h\in
\mathcal{H}:m(X,h)=0$ a.s.$\}.$ By Assumption A3, $\mathcal{H}_{0}$ is
non-empty. The penalty function $P(h)=||h||^{2}$ is strictly convex and
continuous and $||m(\cdot;h)||^{2}$ is convex and continuous. Their Assumption
3.1(i) trivially holds since $W=I_{p}.$ Their Assumption 3.1(iii) is A3(i-ii).
Their Assumption 3.1(iv) follows from A3(ii) since%
\[
||m(\cdot;\Pi_{n}(h_{0}))||^{2}\leq\left\Vert \Pi_{n}(h_{0})-h_{0}\right\Vert
^{2}=o(1).
\]
To verify their Assumption 3.2(c) we need to check%
\begin{equation}
\sup_{h\in\mathcal{H}_{n}}\left\vert \left\Vert h\right\Vert _{n}%
^{2}-\left\Vert h\right\Vert ^{2}\right\vert =o_{P}(1) \label{UP}%
\end{equation}
and%
\[
\left\vert \left\Vert \Pi_{n}(h_{0})\right\Vert ^{2}-\left\Vert h_{0}%
\right\Vert ^{2}\right\vert =o(1).
\]
The last equality follows because $\left\vert \left\Vert \Pi_{n}%
(h_{0})\right\Vert ^{2}-\left\Vert h_{0}\right\Vert ^{2}\right\vert \leq
C\left\Vert \Pi_{n}(h_{0})-h_{0}\right\Vert =o(1).$ Condition (\ref{UP}) is
our Assumption A3(iii). Assumption 3.3 in \cite{Chen_Pouzo_2012} follows from
their Lemma C.2 and our Assumption A2. Assumption 3.4 in
\cite{Chen_Pouzo_2012} is satisfied for the $L_{2}$ norm. Finally, Assumption
A3(iv) completes the conditions of Theorem A.1 in \cite{Chen_Pouzo_2012}, and
hence implies that $\left\Vert \hat{h}_{n}-h_{0}\right\Vert =o_{P}(1).$
$\blacksquare$\bigskip

\noindent\textbf{Lemma A2}: Let Assumptions A1-A4 hold. Then, $\left\Vert
\hat{h}_{n}-h_{0}\right\Vert =O_{P}(\lambda_{n}^{\min(\alpha_{h},2)}%
+\lambda_{n}^{-1}c_{n})$ and $\left\Vert \hat{g}_{n}-g_{0}\right\Vert
=o_{P}(\lambda_{n}^{\min(\alpha_{g},2)}+\lambda_{n}^{-1}c_{n}).$

\noindent\textbf{Proof of Lemma A2}: For simplicity of exposition we consider
the case $p=q=1.$ The proof for $p>1$ or $q>1$ follows the same steps. By the
triangle inequality, with $h_{\lambda_{n}}$ defined in (\ref{App}),
\[
\left\Vert \hat{h}_{n}-h_{0}\right\Vert \leq\left\Vert \hat{h}_{n}%
-h_{\lambda_{n}}\right\Vert +\left\Vert h_{\lambda_{n}}-h_{0}\right\Vert .
\]
Under $h_{0}\in\mathcal{R}(\left(  T^{\ast}T\right)  ^{\alpha_{h}/2}),$ Lemma
A1(1) in \cite{FJV_2011} yields%
\begin{equation}
\left\Vert h_{\lambda_{n}}-h_{0}\right\Vert =O(\lambda_{n}^{\min(\alpha
_{h},2)}). \label{bias}%
\end{equation}
With some abuse of notation, denote $\hat{A}_{\lambda_{n}}=\hat{T}^{\ast}%
\hat{T}+\lambda_{n}I.$ Then, arguing as in Proposition 3.14 of \cite{CFR_hb},
it is shown that%
\begin{equation}
\hat{h}_{n}-h_{\lambda_{n}}=\hat{A}_{\lambda_{n}}^{-1}\hat{T}^{\ast}(\hat
{X}-\hat{T}h_{0})+\hat{A}_{\lambda_{n}}^{-1}(\hat{T}^{\ast}\hat{T}-T^{\ast
}T)(h_{\lambda_{n}}-h_{0}), \label{exp1}%
\end{equation}
and thus,%
\begin{equation}
\left\Vert \hat{h}_{n}-h_{\lambda_{n}}\right\Vert \leq\left\Vert \hat
{A}_{\lambda_{n}}^{-1}\right\Vert \left\Vert \hat{T}^{\ast}(\hat{X}-\hat
{T}h_{0})\right\Vert +\left\Vert \hat{A}_{\lambda_{n}}^{-1}\right\Vert
\left\Vert \hat{T}^{\ast}\hat{T}-T^{\ast}T\right\Vert \left\Vert
h_{\lambda_{n}}-h_{0}\right\Vert . \label{ine1}%
\end{equation}
As in \cite{CFR_hb},%
\[
\left\Vert \hat{A}_{\lambda_{n}}^{-1}\right\Vert =O_{P}(\lambda_{n}^{-1}).
\]
Since $\hat{T}^{\ast}$ is a bounded operator%
\begin{align*}
\left\Vert \hat{T}^{\ast}(\hat{X}-\hat{T}h_{0})\right\Vert  &  =O_{P}\left(
\left\Vert (\hat{X}-\hat{T}h_{0})\right\Vert \right) \\
&  =O_{P}\left(  c_{n,T}\right)  ,
\end{align*}
where recall $c_{n,T}=K_{n}/n+K_{n}^{-2\alpha_{T}},$ and where the second
equality follows from an application of Theorem 1 in \cite{Newey_1997} with
$y=x-h_{0}(z)$ there. Note that Assumption 3 and Assumption A2(iv) imply that
$Var[\left.  y\right\vert X]$ is bounded (which is required in Assumption 1 in
\cite{Newey_1997}). Also note that the supremum bound in Assumption 3 in
\cite{Newey_1997} can be replaced by our $L_{2}-$bound in Assumption A2(iii)
when the goal is to obtain $L_{2}-$rates.

On the other hand,
\begin{equation}
\left\Vert \hat{T}^{\ast}\hat{T}-T^{\ast}T\right\Vert \leq O_{P}\left(
\left\Vert \hat{T}^{\ast}-T^{\ast}\right\Vert \right)  +O_{P}\left(
\left\Vert \hat{T}-T\right\Vert \right)  \label{ine2}%
\end{equation}
and%
\begin{align}
\left\Vert \hat{T}^{\ast}-T^{\ast}\right\Vert  &  \leq\left\Vert
P_{\mathcal{H}_{n}}\right\Vert \left\Vert \hat{T}-T\right\Vert +\left\Vert
P_{\mathcal{H}_{n}}-T^{\ast}\right\Vert \nonumber\\
&  =O_{P}\left(  \left\Vert \hat{T}-T\right\Vert \right)  +O_{P}(c_{n,T^{\ast
}}). \label{ine3}%
\end{align}
We now proceed to establish rates for $\left\Vert \hat{T}-T\right\Vert .$ As
in \cite{Newey_1997}, we can assume without loss of generality that
$E[q^{J_{n}}(Z)q^{J_{n}}(Z)^{\prime}]$ is the identity matrix. Then, by the
triangle inequality,%
\begin{align*}
\left\Vert \hat{T}-T\right\Vert  &  =\sup_{h\in\mathcal{H},\left\Vert
h\right\Vert \leq1}\left\Vert \hat{T}h-Th\right\Vert \\
&  \leq\sup_{h\in\mathcal{H},\left\Vert h\right\Vert \leq1}\left\Vert \hat
{T}h-\pi_{n,p}(h)p^{K_{n}}(\cdot)\right\Vert +\sup_{h\in\mathcal{H},\left\Vert
h\right\Vert \leq1}\left\Vert E[\left.  h(Z)\right\vert X=\cdot]-\pi
_{n,p}(h)p^{K_{n}}(\cdot)\right\Vert \\
&  \leq\sup_{h\in\mathcal{H},\left\Vert h\right\Vert \leq1}\left\Vert \hat
{\pi}_{n,p}(h)-\pi_{n,p}(h)\right\Vert +O(K_{n}^{-\alpha_{T}}),
\end{align*}
where
\[
\hat{\pi}_{n,p}(h)=(P^{\prime}P)^{-1}\sum_{i=1}^{n}p^{K_{n}}(X_{i})h(Z_{i}).
\]
Write%
\[
\hat{\pi}_{n,p}(h)-\pi_{n,p}(h)=Q_{2n}^{-1}P^{\prime}\varepsilon_{h}%
/n+Q_{2n}^{-1}P^{\prime}(G_{h}-P\pi_{n,p}(h))/n,
\]
where $\varepsilon_{h}=H-G_{h},$ $H=(h(Z_{1}),...,h(Z_{n}))^{\prime},$ and
$G_{h}=(Th(X_{1}),...,Th(X_{n}))^{\prime}.$ Similarly to the proof of Theorem
1 in \cite{Newey_1997}, it is shown that%
\[
\sup_{h\in\mathcal{H},\left\Vert h\right\Vert \leq1}\left\Vert Q_{2n}%
^{-1}P^{\prime}\varepsilon_{h}/n\right\Vert ^{2}=O_{P}(K_{n}/n),
\]
where we use Assumption A2(iv) to show that%
\[
\sup_{h\in\mathcal{H},\left\Vert h\right\Vert \leq1}E[\left.  \varepsilon
_{h}\varepsilon_{h}^{\prime}\right\vert X]\leq CI_{n}.
\]
That is,
\begin{align*}
\sup_{h\in\mathcal{H},\left\Vert h\right\Vert \leq1}E\left[  \left.
\left\vert Q_{2n}^{-1/2}P^{\prime}\varepsilon_{h}/n\right\vert ^{2}\right\vert
X\right]   &  =\sup_{h\in\mathcal{H},\left\Vert h\right\Vert \leq1}E\left[
\left.  \varepsilon_{h}P(P^{\prime}P)^{-1}P^{\prime}\varepsilon_{h}\right\vert
X\right]  /n\\
&  =\sup_{h\in\mathcal{H},\left\Vert h\right\Vert \leq1}E\left[  \left.
tr\{P(P^{\prime}P)^{-1}P^{\prime}\varepsilon_{h}\varepsilon_{h}^{\prime
}\}\right\vert X\right]  /n\\
&  =\sup_{h\in\mathcal{H},\left\Vert h\right\Vert \leq1}tr\{P(P^{\prime
}P)^{-1}P^{\prime}E[\left.  \varepsilon_{h}\varepsilon_{h}^{\prime}\right\vert
X]\}/n\\
&  \leq Ctr\{P(P^{\prime}P)^{-1}P^{\prime}\}/n\\
&  \leq CK_{n}/n
\end{align*}
Similarly, by A2(iii)
\[
\sup_{h\in\mathcal{H},\left\Vert h\right\Vert \leq1}\left\Vert Q_{2n}%
^{-1}P^{\prime}(G_{h}-P\pi_{n,p}(h))/n\right\Vert =O_{P}(K_{n}^{-\alpha_{T}%
}).
\]
Then, conclude $\left\Vert \hat{T}-T\right\Vert =O_{P}(c_{n,T})$, $\left\Vert
\hat{T}^{\ast}\hat{T}-T^{\ast}T\right\Vert =O_{P}(c_{n}),$ where
$c_{n}=c_{n,T}+c_{n,T^{\ast}},$ and by (\ref{ine1}), (\ref{ine2}) and
(\ref{ine3})%
\[
\left\Vert \hat{h}_{n}-h_{\lambda_{n}}\right\Vert =O_{P}\left(  \lambda
_{n}^{-1}c_{n}\right)  .
\]
The proof for $\hat{g}_{n}$ is the same and hence omitted. $\blacksquare
$\bigskip

Define the classes%
\[
\mathcal{F}=\{f(y,x,z)=h(z)(y-x^{\prime}\beta_{0}):h\in\mathcal{H}\}.
\]
and%
\[
\mathcal{G}=\{g(y,x,z)=h(z)x:h\in\mathcal{H}\}.
\]

\noindent\textbf{Lemma A3}:

\begin{description}
\item[(i)] Assume $0<E[\left\vert X\right\vert ^{2}]<C$. Then, $N_{[\cdot
]}(\epsilon,\mathcal{G},\left\Vert \cdot\right\Vert _{1})\leq N_{[\cdot
]}(\epsilon/\left\Vert X\right\Vert _{2},\mathcal{H},\left\Vert \cdot
\right\Vert _{2}).$

\item[(ii)] Assume $Var[\left.  Y-X^{\prime}\beta_{0}\right\vert Z]$ is
bounded. Then, $J_{[\cdot]}(\delta,\mathcal{F},\left\Vert \cdot\right\Vert
)<\infty$ if $J_{[\cdot]}(\delta,\mathcal{H},\left\Vert \cdot\right\Vert
)<\infty$ for some $\delta>0.$

\item[(iii)] $N_{[\cdot]}(\epsilon,\mathcal{H}{\footnotesize \cdot}%
\mathcal{G},\left\Vert \cdot\right\Vert _{1})\leq N_{[\cdot]}(C\epsilon
,\mathcal{H},\left\Vert \cdot\right\Vert _{2})\times N_{[\cdot]}%
(C\epsilon,\mathcal{G},\left\Vert \cdot\right\Vert _{2}).$
\end{description}

\noindent\textbf{Proof of Lemma A3}: (i) Let $[l_{j}(Z)X,u_{j}(Z)X]$ be an
$\epsilon/E[\left\vert x\right\vert ^{2}]$ bracket for $\mathcal{H}$. Then, by
Cauchy-Schwartz inequality%
\begin{align*}
\left\Vert l_{j}(Z)X-u_{j}(Z)X\right\Vert _{1}  &  \leq\left\Vert
l_{j}(Z)-u_{j}(Z)\right\Vert \left\Vert X\right\Vert \\
&  \leq\epsilon.
\end{align*}
This shows (i)$.$ The proof of (ii) is analogous, and follows from
\[
\left\Vert l_{j}(Z)U-u_{j}(Z)U\right\Vert \leq C\left\Vert l_{j}%
(Z)-u_{j}(Z)\right\Vert \leq C\epsilon,
\]
where $C$ is such that $Var[\left.  Y-X^{\prime}\beta_{0}\right\vert Z]<C$
a.s. The proof of (iii) is standard and hence omitted. $\blacksquare$

\section{Appendix B: Proofs of Main Results}

\label{AppendixB}

\noindent\textbf{Proof of Lemma \ref{Lemma1}}: The $n^{1/2}$-estimability of
the OLIVA implies the $n^{1/2}$-estimability of the vector-valued functional
\[
E[Xg(X)],
\]
which in turn implies that of the functional
\[
E[X_{j}g(X)],
\]
for each component $X_{j}$ of $X$ (i.e. $X=(X_{1},...,X_{p})^{\prime}).$ By
Lemma 4.1 in \cite{Severini_Tripathi_2012}, the latter implies existence of
$h_{j}\in L_{2}(Z)$ such that%
\[
E[\left.  h_{j}(Z)\right\vert X]=X_{j}\text{ a.s.}%
\]
This implies Assumption 3 with $h(Z)=(h_{1}(Z),...,h_{p}(Z))^{\prime}$.
$\blacksquare$\bigskip

\noindent\textbf{Proof of Proposition \ref{Proposition 1}}: We shall show that
for any $h(Z)\in L_{2}(Z)$ such that%
\[
E[\left.  h(Z)\right\vert X]=X\text{ a.s.}%
\]
the parameter $\beta=E[h(Z)X^{\prime}]^{-1}E[h(Z)Y]$ is uniquely defined.
First, it is straightforward to show that for any such $h,$ $E[h(Z)X^{\prime
}]^{-1}=E[XX^{\prime}]^{-1}.$ Second, we can substitute $Y=g_{0}%
(X)+P_{\mathcal{N}}g(X)+\varepsilon,$ where recall $\mathcal{N}\equiv\{f\in
L_{2}(X):T^{\ast}f=0\}$ and $T^{\ast}f(z):=E[\left.  f(X)\right\vert Z=z].$
Note that for all $h$, $E[h(Z)P_{\mathcal{N}}g(X)]=0,$ so that%
\begin{align*}
E[h(Z)Y]  &  =E[h(Z)g_{0}(X)]\\
&  =E[Xg_{0}(X)],
\end{align*}
for all $h$ satisfying $E[\left.  h(Z)\right\vert X]=X$ a.s. $\blacksquare
$\bigskip

\noindent\textbf{Proof of Proposition \ref{Proposition 2}}: We shall show that
under the conditions of the proposition there exists a $h(Z)\in L_{2}(Z)$ such
that%
\[
E[\left.  h(Z)\right\vert X]=X\text{ a.s.}%
\]
Denote $\bar{\pi}=E[\pi(Z)].$ For a binary $X,$ and since $0<\bar{\pi}<1,$ the
last display is equivalent to the system
\[
E[Xh(Z)]=\bar{\pi}\text{ and }E[(1-X)h(Z)]=0,
\]
or
\[
E[h(Z)]=\bar{\pi}\text{ and }E[\pi(Z)h(Z)]=\bar{\pi}.
\]
Each equation from the last display defines a hyperplane in $h.$ Since
$\pi(Z)$ is not constant, the normal vectors $1$ and $\pi(Z)$ are linearly
independent (not proportional). Hence, the two hyperplanes have an non-empty
intersection, showing that there is at least one $h$ satisfying $E[\left.
h(Z)\right\vert X]=X$ a.s.

Moreover, by Theorem 2, pg. 65, in \cite{Luenberger_Book} the minimum norm
solution is the linear combination of $1$ and $\pi(Z)$ that satisfies the
linear constraints, that is, $h_{0}(Z)=\alpha+\gamma\pi(Z)$ such that $\alpha$
and $\gamma$ satisfy the $2\times2$ system%
\[
\left\{
\begin{tabular}
[c]{l}%
$\alpha+\gamma\bar{\pi}=\bar{\pi}$\\
$\alpha\bar{\pi}+\gamma E[\pi^{2}(Z)]=\bar{\pi}.$%
\end{tabular}
\ \ \ \ \ \right.
\]
Note that this system has a unique solution, since the determinant of the
coefficient matrix is $Var(\pi(Z))>0.$ Then, the unique solution is given by
\begin{align*}
\left[
\begin{array}
[c]{c}%
\alpha\\
\gamma
\end{array}
\right]   &  =\left[
\begin{array}
[c]{cc}%
1 & \bar{\pi}\\
\bar{\pi} & E[\pi^{2}(Z)]
\end{array}
\right]  ^{-1}\left[
\begin{array}
[c]{c}%
\bar{\pi}\\
\bar{\pi}%
\end{array}
\right] \\
&  =\left[
\begin{array}
[c]{c}%
\bar{\pi}\left(  1-\frac{\bar{\pi}(1-\bar{\pi})}{var(\pi(Z))}\right) \\
\frac{\bar{\pi}(1-\bar{\pi})}{var(\pi(Z))}%
\end{array}
\right]  .
\end{align*}
$\blacksquare$\bigskip

\noindent\textbf{Proof of Proposition \ref{Proposition 3}}: Assume without
loss of generality that $X$ is scalar and note that, by \cite{EHN},
$h_{1}(Z)=h_{0}(Z)+h_{\perp}(Z),$ with $Cov(h_{0}(Z),h_{\perp}(Z))=0$ (note
$E[h_{0}(Z)h_{\perp}(Z)]=0$ and $E[h_{\perp}(Z)]=0).$ Thus, since $E[\left.
h_{0}(Z)\right\vert X]=X$ and $E[\left.  h_{1}(Z)\right\vert X]=X,$ then
$E[\left.  h_{\perp}(Z)\right\vert X]=0$ a.s., and hence%
\[
0=Cov(X,h_{\perp}(Z))=\alpha_{1}Var(h_{\perp}(Z)),
\]
and hence, if $h_{1}\neq h_{0}$ (i.e. $Var(h_{\perp}(Z))>0$) then $\alpha
_{1}=0.$ $\blacksquare$\bigskip

\noindent\textbf{Proof of Theorem \ref{ANTSIV}}: Write
\begin{align*}
\hat{\beta}  &  =\left(  E_{n}\left[  \hat{h}_{n}(Z_{i})X_{i}^{\prime}\right]
\right)  ^{-1}\left(  E_{n}\left[  \hat{h}_{n}(Z_{i})Y_{i}\right]  \right) \\
&  =\beta_{0}+\left(  E_{n}\left[  \hat{h}_{n}(Z_{i})X_{i}^{\prime}\right]
\right)  ^{-1}\left(  E_{n}\left[  \hat{h}_{n}(Z_{i})U_{i}\right]  \right)  .
\end{align*}
Note that%
\begin{align}
E_{n}\left[  \hat{h}_{n}(Z_{i})X_{i}^{\prime}\right]   &  =E_{n}\left[
h_{0}(Z_{i})X_{i}^{\prime}\right]  +o_{P}(1)\nonumber\\
&  =E\left[  h_{0}(Z_{i})X_{i}^{\prime}\right]  +o_{P}(1), \label{exp9}%
\end{align}
where the first equality follows from Lemma A3(i), Lemma A1, Assumption A5 and
$\hat{h}_{n}\in\mathcal{H}$ by an application of a Glivenko-Cantelli%
\'{}%
s argument, and the second equality follows from the Law of Large Numbers. The
same arguments show that $E_{n}\left[  \hat{h}_{n}(Z_{i})U_{i}\right]
=o_{P}(1).$ Thus, $\hat{\beta}$ is consistent for $\beta_{0}.$

Likewise, Lemma A3(ii), Lemma A1, Assumption A5(ii) and $\hat{h}_{n}%
\in\mathcal{H},$ yields for $\hat{f}=\hat{h}_{n}(Z_{i})U_{i}$ and $f_{0}%
=h_{0}(Z_{i})U_{i},$%
\[
\mathbb{G}_{n}\hat{f}=\mathbb{G}_{n}f_{0}+o_{P}(1),
\]
since the class $\mathcal{F}$ is a Donsker class, see Theorem 2.5.6 in
\cite{van_der_Vaart_Wellner_Book}. Then,%
\begin{equation}
\sqrt{n}\left(  \hat{\beta}-\beta_{0}\right)  =\left(  E\left[  h_{0}%
(Z_{i})X_{i}^{\prime}\right]  +o_{P}(1)\right)  ^{-1}\left(  \sqrt{n}%
E_{n}\left[  h_{0}(Z_{i})U_{i}\right]  +\sqrt{n}\mathbb{P}\left[  \left\{
\hat{h}_{n}(Z_{i})-h_{0}(Z_{i})\right\}  U_{i}\right]  \right)  . \label{exp3}%
\end{equation}
We investigate the second term, which with the notation $\left\langle
h_{1},h_{2}\right\rangle =E[h_{1}(Z)h_{2}(Z)]$ can be written as%
\[
\sqrt{n}\mathbb{P}\left[  \left\{  \hat{h}_{n}(Z_{i})-h_{0}(Z_{i})\right\}
U_{i}\right]  =\sqrt{n}\left\langle \hat{h}_{n}-h_{0},u\right\rangle
\]
where $u(z)=E[\left.  U\right\vert Z=z]$ is in $L_{2}(Z)$ by A5(i).

From the proof of Lemma A2, and in particular (\ref{bias}) and (\ref{exp1}),
and Assumption A6(ii),
\begin{align*}
\sqrt{n}\left\langle \hat{h}_{n}-h_{0},u\right\rangle  &  =\sqrt
{n}\left\langle \hat{h}_{n}-h_{\lambda_{n}},u\right\rangle +\sqrt
{n}\left\langle h_{\lambda_{n}}-h_{0},u\right\rangle \\
&  =\sqrt{n}\left\langle \hat{A}_{\lambda_{n}}^{-1}\hat{T}^{\ast}(\hat{X}%
-\hat{T}h_{0}),u\right\rangle +O_{P}\left(  \sqrt{n}c_{n}\lambda_{n}%
^{\min(\alpha_{h}-1,1)}\right)  +O\left(  \sqrt{n}\lambda_{n}^{\min(\alpha
_{h},2)}\right) \\
&  =\sqrt{n}\left\langle \hat{A}_{\lambda_{n}}^{-1}\hat{T}^{\ast}(\hat{X}%
-\hat{T}h_{0}),u\right\rangle +o_{P}\left(  1\right)  .
\end{align*}
Next, we write
\begin{align*}
\sqrt{n}\left\langle \hat{A}_{\lambda_{n}}^{-1}\hat{T}^{\ast}(\hat{X}-\hat
{T}h_{0}),u\right\rangle  &  =\sqrt{n}\left\langle A_{\lambda_{n}}^{-1}%
T^{\ast}(\hat{X}-\hat{T}h_{0}),u\right\rangle \\
&  +\sqrt{n}\left\langle \left(  \hat{A}_{\lambda_{n}}^{-1}-A_{\lambda_{n}%
}^{-1}\right)  T^{\ast}(\hat{X}-\hat{T}h_{0}),u\right\rangle \\
&  +\sqrt{n}\left\langle A_{\lambda_{n}}^{-1}\left(  \hat{T}^{\ast}-T^{\ast
}\right)  (\hat{T}X-\hat{T}h_{0}),u\right\rangle \\
&  +\sqrt{n}\left\langle \left(  \hat{A}_{\lambda_{n}}^{-1}-A_{\lambda_{n}%
}^{-1}\right)  \left(  \hat{T}^{\ast}-T^{\ast}\right)  (\hat{X}-\hat{T}%
h_{0}),u\right\rangle \\
&  \equiv C_{1n}+C_{2n}+C_{3n}+C_{4n}.
\end{align*}
From the simple equality $B^{-1}-C^{-1}=B^{-1}(C-B)C^{-1}$ we obtain $\hat
{A}_{\lambda_{n}}^{-1}-A_{\lambda_{n}}^{-1}=\hat{A}_{\lambda_{n}}^{-1}\left(
T^{\ast}T-\hat{T}^{\ast}\hat{T}\right)  A_{\lambda_{n}}^{-1},$ and from this
and Lemma A2,%
\begin{align*}
\left\vert C_{4n}\right\vert  &  =O_{P}(\sqrt{n}\lambda_{n}^{-2}c_{n}%
^{3})=o_{P}(1)\text{, by A6(i);}\\
\left\vert C_{3n}\right\vert  &  =O_{P}(\sqrt{n}\lambda_{n}^{-1}c_{n}%
^{2})=o_{P}(1)\text{, by A6(i);}\\
\left\vert C_{2n}\right\vert  &  =O_{P}(\sqrt{n}\lambda_{n}^{-2}c_{n}%
^{2})=o_{P}(1)\text{, by A6(i).}%
\end{align*}
To analyze the term $C_{1n}$ we use Theorem 3 in \cite{Newey_1997} after
writing
\[
C_{1n}=\sqrt{n}\left\langle \hat{T}\varphi,v_{n}\right\rangle ,
\]
where $\varphi=X-h_{0}$ and
\begin{equation}
v_{n}=TA_{\lambda_{n}}^{-1}u. \label{vn}%
\end{equation}
Note that
\[
u=E[\left.  Y-\beta_{0}^{\prime}X\right\vert Z]=E[\left.  g_{0}(X)-\beta
_{0}^{\prime}X\right\vert Z],
\]
and hence $v_{n}=TA_{\lambda_{n}}^{-1}T^{\ast}\Delta,$ for $\Delta
(X)=(g_{0}(X)-\beta_{0}^{\prime}X).$

Assumption A6(iii) implies Assumptions 1 and 4 in \cite{Newey_1997}.
Assumption A2 implies Assumptions 2 and 3 in \cite{Newey_1997} (with $d=0$
there). Note that by Lemma A1(A.4) in \cite{FJV_2011}%
\[
\left\Vert v_{n}\right\Vert \leq\left\Vert TA_{\lambda_{n}}^{-1}T^{\ast
}\right\Vert \left\Vert \Delta\right\Vert \leq\left\Vert \Delta\right\Vert
<\infty.
\]
Hence, Assumption 7 in \cite{Newey_1997} holds with $g_{0}=T\varphi$ there.
Hence, Theorem 4 in \cite{Newey_1997} applies to $C_{1n}$ to conclude from its
proof that%
\begin{equation}
C_{1n}=-\frac{1}{\sqrt{n}}\sum_{i}^{n}v_{n}(X_{i})(h_{0}(Z_{i})-X_{i}%
)+o_{P}(1). \label{exp4}%
\end{equation}
We will use that $g_{0}(X)-\beta_{0}^{\prime}X$ is in $\mathcal{R}(\left(
TT^{\ast}\right)  ^{\alpha_{g}/2}),$ $\alpha_{g}>0$. Note the identities%
\[
T\left(  T^{\ast}T+\lambda_{n}I\right)  ^{-1}T^{\ast}=\left(  TT^{\ast
}+\lambda_{n}I\right)  ^{-1}TT^{\ast}%
\]
and%
\[
I-\left(  TT^{\ast}+\lambda_{n}I\right)  ^{-1}TT^{\ast}=\lambda_{n}\left(
TT^{\ast}+\lambda_{n}I\right)  ^{-1}.
\]
Then,
\begin{equation}
\frac{1}{\sqrt{n}}\sum_{i}^{n}v_{n}(X_{i})(h_{0}(Z_{i})-X_{i})=\frac{1}%
{\sqrt{n}}\sum_{i}^{n}\left(  g_{0}(X_{i})-\beta_{0}^{\prime}X_{i}\right)
(h_{0}(Z_{i})-X_{i})+o_{P}(1), \label{exp5}%
\end{equation}
since by Lemma A1(A1) in \cite{FJV_2011}, Assumption A4(i) and Assumption 3,
it holds%
\begin{align*}
Var\left(  \frac{1}{\sqrt{n}}\sum_{i}^{n}\left[  v_{n}(X_{i})-\left(
g_{0}(X_{i})-\beta_{0}^{\prime}X_{i}\right)  \right]  (h_{0}(Z_{i}%
)-X_{i})\right)   &  \leq C\left\Vert v_{n}(X_{i})-\left(  g_{0}(X_{i}%
)-\beta_{0}^{\prime}X_{i}\right)  \right\Vert ^{2}\\
&  =C\left\Vert \lambda_{n}\left(  TT^{\ast}+\lambda_{n}I\right)  ^{-1}\left(
g_{0}(X_{i})-\beta_{0}^{\prime}X_{i}\right)  \right\Vert ^{2}\\
&  \leq C\lambda_{n}^{\min(\alpha_{g},2)}.
\end{align*}
Thus, from (\ref{exp3}), (\ref{exp4}) and (\ref{exp5})
\[
\sqrt{n}\left(  \hat{\beta}-\beta_{0}\right)  =\left(  E\left[  h_{0}%
(Z_{i})X_{i}^{\prime}\right]  \right)  ^{-1}\sqrt{n}E_{n}\left[  m(W_{i}%
,\beta_{0},h_{0},g_{0})\right]  +o_{P}(1).
\]
The asymptotic normality then follows from the standard Central Limit Theorem.

We now show the consistency of $\hat{\Sigma}=E_{n}[\hat{h}_{n}(Z_{i}%
)X_{i}^{\prime}]^{-1}E_{n}[\hat{m}_{ni}\hat{m}_{ni}^{\prime}]E_{n}[\hat{h}%
_{n}(Z_{i})X_{i}^{\prime}]^{-1}.$ Write, with $m_{0i}=m(W_{i},\beta
,h_{0},g_{0}),$%
\begin{equation}
E_{n}[\hat{m}_{ni}\hat{m}_{ni}^{\prime}]-E_{n}[m_{0i}m_{0i}^{\prime}%
]=E_{n}[m_{0i}(\hat{m}_{ni}^{\prime}-m_{0i}^{\prime})]+E_{n}[(\hat{m}%
_{ni}-m_{0i})m_{0i}^{\prime}]+E_{n}[(\hat{m}_{ni}-m_{0i})(\hat{m}_{ni}%
-m_{0i})^{\prime}] \label{exp6}%
\end{equation}
and
\[
\hat{m}_{ni}-m_{0i}=\left(  Y-g_{0}(X_{i}\right)  )\left(  \hat{h}_{n}%
(Z_{i})-h_{0}(Z_{i})\right)  -\left(  \hat{g}_{n}(X_{i})-g_{0}(X_{i})\right)
\left(  \hat{h}_{n}(Z_{i})-X_{i}\right)  .
\]
By Cauchy-Schwartz inequality and Assumption 2%
\[
\left\vert E_{n}\left[  m_{0i}\left(  Y-g_{0}(X_{i}\right)  )\left(  \hat
{h}_{n}(Z_{i})-h_{0}(Z_{i})\right)  ^{\prime}\right]  \right\vert ^{2}\leq
CE_{n}\left[  \left\vert \hat{h}_{n}(Z_{i})-h_{0}(Z_{i})\right\vert
^{2}\right]  .
\]
The class of functions%
\[
\{\left\vert h(z)-h_{0}\right\vert ^{2}:h\in\mathcal{H}\}
\]
is Glivenko-Cantelli under the conditions on $\mathcal{H}$, and thus
$E_{n}\left[  \left\vert \hat{h}_{n}(Z_{i})-h_{0}(Z_{i})\right\vert
^{2}\right]  =o_{P}(1)$ by Lemma A1. Likewise,
\begin{align*}
\left\vert E_{n}\left[  m_{0i}^{\prime}\left(  \hat{g}_{n}(X_{i})-g_{0}%
(X_{i})\right)  \left(  \hat{h}_{n}(Z_{i})-X_{i}\right)  ^{\prime}\right]
\right\vert ^{2}  &  \leq CE_{n}\left[  \left\vert \hat{g}_{n}(X_{i}%
)-g_{0}(X_{i})\right\vert ^{2}\right] \\
&  =o_{P}(1),
\end{align*}
by Assumption A5(ii) and Lemma A1. Other terms in (\ref{exp6}) are analyzed
similarly, to conclude that they are $o_{P}(1).$ Together with (\ref{exp9}),
this implies the consistency of $\hat{\Sigma}.$ $\blacksquare$\bigskip

\noindent\textbf{Proof of Theorem \ref{Hausman}}: We first show that the OLS
first-stage estimator $\hat{\alpha}=(\hat{\alpha}_{1}^{\prime},\hat{\alpha
}_{2})^{\prime}$ of $\alpha_{0}=(\alpha_{1}^{\prime},\alpha_{2})^{\prime}$ in
the regression%
\[
X_{2}=\alpha_{1}^{\prime}X_{1}+\alpha_{2}\hat{h}_{2n}(Z)+e,
\]
satisfies $\sqrt{n}(\hat{\alpha}-\alpha_{0})=O_{P}(1)$. Note $e=V-\alpha
_{2}(\hat{h}_{2n}(Z)-h_{20}(Z))$, and denote $\hat{h}_{n}(Z)=(X_{1}^{\prime
},\hat{h}_{2n}(Z))^{\prime}$ and $h_{0}(Z)=(X_{1}^{\prime},h_{20}(Z))^{\prime
}.$ Then,
\[
\sqrt{n}(\hat{\alpha}-\alpha_{0})=\left(  E_{n}\left[  \hat{h}_{n}^{\prime
}\hat{h}_{n}^{\prime}\right]  \right)  ^{-1}\sqrt{n}E_{n}\left[  \hat{h}%
_{n}e\right]  .
\]
Lemma A2 and a Glivenko-Cantelli%
\'{}%
s argument imply $E_{n}\left[  \hat{h}_{n}\hat{h}_{n}^{\prime}\right]
=E_{n}\left[  h_{0}(Z)h_{0}^{\prime}(Z)\right]  +o_{P}(1)=O_{P}(1).$

By $\left\Vert \hat{h}_{2n}-h_{20}\right\Vert =o_{P}(n^{-1/4}),$ it holds%
\begin{align*}
\sqrt{n}E_{n}\left[  \hat{h}_{n}(Z)e\right]   &  =\sqrt{n}E_{n}\left[  \hat
{h}_{n}(Z)V\right]  -\alpha_{2}\sqrt{n}E_{n}\left[  \hat{h}_{n}(Z)(\hat
{h}_{2n}(Z)-h_{20}(Z))\right] \\
&  =\sqrt{n}E_{n}\left[  h_{0}(Z)V\right]  -\alpha_{2}\sqrt{n}E_{n}\left[
h_{0}(Z)(\hat{h}_{2n}(Z)-h_{20}(Z))\right]  +\sqrt{n}E_{n}\left[  (\hat{h}%
_{n}(Z)-h_{0}(Z))V\right]  +o_{P}(1)\\
&  \equiv A_{1}-\alpha_{2}A_{2}+A_{3}+o_{P}(1).
\end{align*}
The standard central limit theorem implies $A_{1}=O_{P}(1).$

An empirical processes argument shows%
\[
A_{2}=\sqrt{n}E\left[  h_{0}(Z)(\hat{h}_{2n}(Z)-h_{20}(Z))\right]  +o_{P}(1).
\]
By A6(ii),%
\begin{align*}
\sqrt{n}E\left[  h_{0}(Z)(\hat{h}_{2n}(Z)-h_{20}(Z))\right]   &  =\sqrt
{n}E\left[  h_{0}(Z)(\hat{h}_{2n}(Z)-h_{\lambda_{n}}(Z))\right]  +\sqrt
{n}E\left[  h_{0}(Z)(h_{\lambda_{n}}(Z)-h_{20}(Z))\right] \\
&  =\sqrt{n}E\left[  h_{0}(Z)(\hat{h}_{2n}(Z)-h_{\lambda_{n}}(Z))\right]
+o_{P}(1).
\end{align*}
While (\ref{exp1}) and A6(ii) yield
\begin{align*}
A_{2}  &  =\sqrt{n}E\left[  h_{0}(Z)\hat{A}_{\lambda_{n}}^{-1}\hat{T}^{\ast
}(\hat{X}-\hat{T}h_{0})(Z)\right]  +o_{P}(1)\\
&  =\sqrt{n}E\left[  h_{0}(Z)A_{\lambda_{n}}^{-1}T^{\ast}(\hat{X}-\hat{T}%
h_{0})(Z)\right]  +o_{P}(1)\\
&  \equiv\sqrt{n}E\left[  v(Z)(\hat{X}-\hat{T}h_{0})(Z)\right]  +o_{P}(1),
\end{align*}
where $v(Z)=TA_{\lambda_{n}}^{-1}h_{0}(Z).$ By $h_{0}\in\mathcal{R}(T^{\ast
}),$ $h_{0}=T^{\ast}\psi$ for some $\psi$ with $\left\Vert \psi\right\Vert
<\infty,$ then by Lemma A1(A.4) in \cite{FJV_2011}
\begin{align*}
\left\Vert v\right\Vert  &  \leq\left\Vert TA_{\lambda_{n}}^{-1}T^{\ast
}\right\Vert \left\Vert \psi\right\Vert \\
&  \leq\left\Vert \psi\right\Vert <\infty.
\end{align*}
Then, by Theorem 3 in \cite{Newey_1997}, $A_{2}=O_{P}(1).$ A similar argument
as for $A_{2}$ shows $A_{3}=O_{P}(1),$ because $E[\left.  V\right\vert
Z]\in\mathcal{R}(T^{\ast}).$ Thus, combining the previous bounds we obtain
$\sqrt{n}(\hat{\alpha}-\alpha_{0})=O_{P}(1)$.

We proceed now with second step estimator. Denote $\hat{S}=(X,\hat{V}%
)^{\prime}$ and $\theta=(\beta^{\prime},\rho)^{\prime}.$ Let $\hat{\theta}$
denote the OLS of $Y$ on $\hat{S}.$ Since, since under the null $\rho=0,$ then%
\begin{align*}
\hat{\theta}  &  =\left(  E_{n}\left[  \hat{S}\hat{S}^{\prime}\right]
\right)  ^{-1}E_{n}\left[  \hat{S}Y\right] \\
&  =\theta+\left(  E_{n}\left[  \hat{S}\hat{S}^{\prime}\right]  \right)
^{-1}E_{n}\left[  \hat{S}U\right] \\
&  =\theta+\left(  E\left[  SS^{\prime}\right]  \right)  ^{-1}E_{n}\left[
SU\right]  +\left(  E\left[  SS^{\prime}\right]  \right)  ^{-1}E_{n}\left[
(\hat{S}-S)U\right]  +o_{P}(n^{-1/2})\\
&  =\theta+\left(  E\left[  SS^{\prime}\right]  \right)  ^{-1}E_{n}\left[
SU\right]  +o_{P}(n^{-1/2}),
\end{align*}
where the last equality follows because%
\begin{align*}
\sqrt{n}E_{n}\left[  (\hat{V}-V)U\right]   &  =\sqrt{n}(\hat{\alpha}%
-\alpha_{0})^{\prime}E_{n}\left[  h_{0}(Z)U\right]  +\hat{\alpha}_{2}\sqrt
{n}E_{n}\left[  U(\hat{h}_{2n}(Z)-h_{20}(Z))\right] \\
&  =O_{P}(1)\times o_{P}(1)+O_{P}(1)\times o_{P}(1),
\end{align*}
with the term $\sqrt{n}E_{n}\left[  U(\hat{h}_{2n}(Z)-h_{20}(Z))\right]  $
being $o_{P}(1)$ because by A6(iv)%
\begin{align*}
\sqrt{n}E_{n}\left[  U(\hat{h}_{2n}(Z)-h_{20}(Z))\right]   &  =\sqrt
{n}\mathbb{P}\left[  U(\hat{h}_{2n}(Z)-h_{20}(Z))\right]  +o_{P}(1)\\
&  =o_{P}(1).
\end{align*}
Thus, the standard asymptotic normality for the OLS estimator applies.
$\blacksquare$\newpage%
%

\end{document}